\documentclass[review]{elsarticle}
\makeatletter
\def\ps@pprintTitle{%
 \let\@oddhead\@empty
 \let\@evenhead\@empty
 \def\@oddfoot{}%
 \let\@evenfoot\@oddfoot}
\makeatother
\usepackage{mathrsfs} 
\usepackage{hyperref,graphicx}
\usepackage[normalem]{ulem}
\usepackage[margin=1.0in]{geometry}
\usepackage{multirow}
\usepackage{subfigure}
\usepackage{physics}
\usepackage{braket}
\usepackage{dsfont}
\usepackage[cmintegrals,varg]{newtxmath} 

\allowdisplaybreaks[2]
\usepackage{amsfonts}
\usepackage{amsmath,bm}
\usepackage{booktabs, threeparttable}
\usepackage{array}
\newcolumntype{L}[1]{>{\raggedright\arraybackslash}p{#1}}
\usepackage{color}

\begin{document}

\title{Thermal response functions and second sound in single-layer hexagonal boron nitride}

\author[ucfaddress,AMPACaddress]{Patrick K. Schelling\corref{cor1}}
\ead{patrick.schelling@ucf.edu}
\author[ucfaddress] {Antonio Martinez Margolles, and Logan P. Echazabal}
\cortext[cor1]{Corresponding author.}
\address[ucfaddress]{Department of Physics, University of Central Florida, Orlando, FL 32816-2385, USA}
\address[AMPACaddress]{Advanced Materials Processing and Analysis Center, University of Central Florida, Orlando, FL 32816-2385, USA}

\begin{abstract}
Ballistic heat transport and second sound propagation in solids is of direct relevance in electronic and energy applications at short length scales and low temperatures. Measurement or calculation of thermal conductivity, which is typically a primary objective, may be of limited usefulness for predicting heat transport which does not follow the heat-diffusion equation. In this paper, molecular-dynamics simulations of hexagonal BN (h-BN) are used
to compute thermal response functions from equilibrium correlation functions defined in Fourier space. The response functions
are useful for describing the time-dependent transport beyond the usual assumptions of Fourier's law. The results demonstrate that for length scales $\sim 110$nm at $T=100$K second sound should be experimentally observable. At higher temperatures and longer length scales, while second sound may not be directly
observable, thermal transport can nevertheless strongly deviate from predictions based on the heat-diffusion equation. Along with classical simulations, we outline a first-principles, many-body theoretical
approach for calculation of the response function based on solutions of the Bethe-Salpeter equation. The relevant expressions for heat current clarify the importance of phase coherence within
a phonon branch to the observation of second sound. Previous work on one-dimensional chains is also discussed to show that materials characterized by
linear dispersion and simple phonon band structure should more readily display second sound.
 \end{abstract}


\maketitle

\section{Introduction}
In recent years there has been significant interest in phonon heat transport in conditions where Fourier's law and diffusive heat transport do not apply. For
example, in conditions where momentum-conserving normal phonon scattering dominates Umklapp scattering, phonon hydrodynamics is expected to be observed\cite{https://doi.org/10.1002/pssa.2210240102,Cepellotti_2015,Ghosh_2022}. This phenomenon
has been discussed especially for two-dimensional materials at low temperatures\cite{Cepellotti_2015}. In addition to phonon-hydrodynamics, low-temperature conditions may result in transport via second sound\cite{Chester:1963vm,Guyer:1966ab,Sham:1967aa,Sham:1967ab,Hardy:1970aa,Jackson:1970aa}. Second sound refers to heat transport that is wave-like and oscillatory. Observations of second sound have been reported in bulk three-dimensional crystals including Bi \cite{Narayanamurti:1972aa} and Ge \cite{beardo2021observation}.  Quite recently, second sound was observed experimentally in graphite using a sub-picosecond transient grating technique, with the observations analyzed using solutions of the Peierls-Boltzmann equation (PBE) \cite{huberman_2019,Ding_2022}. 
Generally, theoretical descriptions of second sound have been based on the hyperbolic heat equation with solutions to the PBE used to relate propagation velocity to the underlying phonon spectra\cite{Hardy:1970aa,Lee:2017aa,Luo:2019aa,Shang:2022aa}. 

However, despite the experimental and theoretical progress in this area, methodology has not yet been developed to elucidate transport resulting from time-dependent heating sources in regimes where Fourier's law does not apply. Moreover, many theoretical calculations and experimental measurements focus on determining values for thermal conductivity, despite the fact that the heat-diffusion equation does not always apply.  Specifically, at shorter length and time scales, ballistic transport and second sound may play a dominant role. We currently do not have theoretical methodology that provides a meaningful description in these regimes. It is evident that there needs to be significant effort to address these shortcomings in our understanding. Finally, the theoretical methodology needed should be able to describe not only ballistic regimes where phonon hydrodynamics and second sound are prevalent, but also at intermediate scales and finally towards conditions where the heat-diffusion equation becomes more reliable.

Recently we have developed an approach which uses classical molecular-dynamics (MD) simulation and current-current correlation functions to compute thermal-response functions \cite{Fernando_2020}. The thermal-response function describes how a system evolves as the result of time-dependent heat inputs. The advantage is that the response is computed without assumptions about the applicability of Fourier's law and can be directly applied to describe time-dependent phenomena. So far the approach has been applied to transport in Lennard-Jones (LJ) solids\cite{Fernando_2020} and in 1D chains \cite{Bohm:2022aa}. The analogous concept ``thermal susceptibility'' has been explored in recent publications by Allen and coworkers\cite{Allen:2022we,Kefayati:2022aa}. Especially in the case of 1D chains, our approach demonstrated the presence of second-sound propagation\cite{Bohm:2022aa} in response to localized heat pulses. However, detailed studies using this approach in realistic two- and three-dimensional materials has not yet been reported.

In this paper we report thermal transport properties determined from equilibrium MD simulations of single-layer hexagonal boron nitride (h-BN). The principal objective of this paper is to elucidate second-sound in h-BN so that experimentalists can have an idea of conditions where it could be observed. Second sound tends to be 
identified when temperature fields exhibit damped oscillations rather than exponential decay towards equilibrium. This is the most definitive and characteristic signature of second sound. However, our work demonstrates that even in the absence of clear oscillatory behavior, transport in h-BN can still be ballistic, wave-like, and deviate strongly from predictions based on Fourier's law. Our results determine conditions for the observation of second sound and, more broadly, strong deviations from Fourier's law.
Finally, we outline a theoretical approach, which builds on work by Sham\cite{Sham:1967aa,Sham:1967ab}, to describe heat transport via second sound. This methodology holds
significant promise for describing time-dependent heat transport starting from first-principles from the ballistic regime where second sound is relevant to transport in diffusive regimes where Fourier's law is a reasonable approximation.

\section{Approach: Classical MD}

Fan and coworkers developed heat current expressions suitable for many-body Tersoff potentials, and applied their equations to Green-Kubo calculations of thermal
conductivity in various materials systems\cite{Fan:2015tj}. We use their definitions for computation of heat-flux density. We use the Tersoff potential for h-BN first reported in Ref. \cite{Lindsay:2011aa}. This potential was also used in the MD thermal conductivity calculations reported in Ref. \cite{Cai:2019aa}.

Beyond computation of $\kappa$, in this paper we also compute thermal response-functions\cite{Fernando_2020}. This approach to thermal-transport modeling was first developed by us and applied to simple Lennard-Jones solids\cite{Fernando_2020}. The idea is to compute the heat-flux density due to an externally-input heat power density $H^{(ext)}(\bm{r}^{\prime},t^{\prime})$. First, assuming that the Cartesian axes are aligned along principle axes of the crystal such that the response function is a diagonal tensor, we have for the $\mu$th component of the heat-flux density $J_{\mu}(\bm{r},t)$
\begin{equation} \label{def1}
J_{\mu}(\bm{r},t)= -{1 \over C_{V}} \int_{-\infty}^{t} dt^{\prime} \int_{\Omega} d^{3} r^{\prime}  K_{\mu \mu} (\bm{r}-\bm{r}^{\prime},t-t^{\prime}) {\partial H^{(ext)}(\bm{r}^{\prime},t^{\prime}) \over \partial r^{\prime}_{\mu}} \text{  ,}
\end{equation}
in which $C_{V}$ is the system heat capacity and $\Omega$ represent the system volume. Note that it is assumed that the Cartesian axes are chosen
to reflect the symmetry of the crystal such that a gradient in the heat input along a Cartesian direction $\mu$ results in a heat current along the same direction. 

We first write this equation in Fourier space, starting with the heat-flux density,
\begin{equation}
\bm{J}(\bm{r},t) = \sum_{\bm{q}}\bm{J}_{\bm{q}}(t) e^{i\bm{q} \cdot \bm{r}} \text{   .}
\end{equation}
In a practical MD calculations using periodic-boundary conditions, the $\bm{q}$ correspond to reciprocal-lattice vectors for the simulation supercell. The response-function
tensor and input heat power can similarly be written in Fourier space and then Eq. \ref{def1} can be written for each $\bm{q}=q \hat{e}_{\mu}$,
\begin{equation} \label{def2}
J_{\bm{q}}(t) = -{i \over c_{V}} q \int_{-\infty}^{t} dt^{\prime} K_{\bm{q}}(t-t^{\prime}) H^{(ext)}_{\bm{q}}(t^{\prime})  \text{   ,}
\end{equation}
in which $c_{V}={C_{V} \over \Omega}$ is the volumetric specific heat capacity.

To obtain an expression which can be used to compute the response-function tensor, we first of all assume an input heat pulse at $t^{\prime}=0$ such that $H^{(ext)}_{\bm{q}}(t^{\prime}) = u^{(ext)}_{\bm{q}}(0) \delta(t^{\prime})$. Then with the assumption that
the system was originally in equilibrium before the input pulse, and that $t >0$, the Fourier component of the heat-flux density is,
\begin{equation} \label{def3}
J_{\bm{q}}(t) = -{i \over c_{V}} q K_{\bm{q}}(t) u^{(ext)}_{\bm{q}}(0)
\end{equation}
Following the arguments in the original article\cite{Fernando_2020}, we can determine the response function by computing the dissipation of thermal fluctuations within the equilibrium ensemble. Then the response function can be determined for a particular $\bm{q}$ vector,
\begin{equation} \label{corrMD}
 K_{\bm{q}}(\tau) =c_{V} {\int_{0}^{\tau} dt \langle J_{\bm{q}}(t) J_{-\bm{q}}(0)\rangle  \over \langle  u_{\bm{q}}(0) u_{-\bm{q}}(0)\rangle}
 \end{equation}

Calculation of the Fourier components of the heat-flux density $J_{\bm{q}}$ requires a local definition at each atomic site. The derivation of the local heat-flux density
$\bm{J}_{i}$ at each atomic site $i$ is presented in Appendix A. It should be noted that the approach used here does not include the so-called ``convective'' terms, which on general
grounds are understood to be unimportant for heat transport in a crystalline system.

In Appendix B,  the response function assuming the validity of Fourier's law is used to derive the standard Green-Kubo expression for $\kappa$ in the limit $\bm{q} \rightarrow 0$. While the response function for the heat-diffusion equation was discussed in Ref. \cite{Fernando_2020}, the direct connection between response functions and the Green-Kubo equations for thermal conductivity was not demonstrated. This should clarify some aspects of thermal response functions.

\section{Results}

We first explore the dependence of transport calculations on equilibration time. In each of the reported calculations, an initial constant-$T$ simulation, using an integration time step of $0.367$fs, was performed for $10^{4}$ MD steps ($3.67$ps), followed by equilibration under constant-energy conditions. The total integration time for equilibration is denoted by $\tau_{eq}$. For time averages used for transport calculations, we launched $64$ independent simulations. Altogether from these simulations, the total integration time used for averaging was $94.08$ns. Each independent simulation included $1.47$ns of averaging time, which is substantially greater than the correlation times. This appears to be in agreement with Fan et al.\cite{Fan:2015tj} for graphene simulations, where convergence at $T=300$K was observed for conductivity calculations by about $250$ps of integration time. As will become apparent, correlation times for response functions with $\bm{q} \ne 0$ tend to converge even more rapidly. All simulations were conducted in constant-volume conditions with periodic-boundary
conditions in two dimensions. We first present results for a rather small system with $N=2048$ atoms. The simulation supercell was taken to be $32\times 32$ primitive cells along the primitive $\bm{a}_{1}$ and $\bm{a}_{2}$ Bravais lattice vectors. The dimensions of the supercell were chosen to result in an h-BN $T=0$K bond length of $1.443 \AA$ and lattice parameter $a_{0}=2.499\AA$. This $T=0$K bond length was used
for all calculations independent of simulation temperature $T$ and system size. Due to the tendency of two-dimensional materials to exhibit thermal contraction with increasing temperature,
the simulated system experienced a fairly small tensile stress. In computing a thermal conductivity, we assume an additional length scale $c=3.33\AA$ which is equal to the layer-spacing in multilayer h-BN.

\begin{figure}
\begin{centering}
\includegraphics[width=0.40\textwidth]{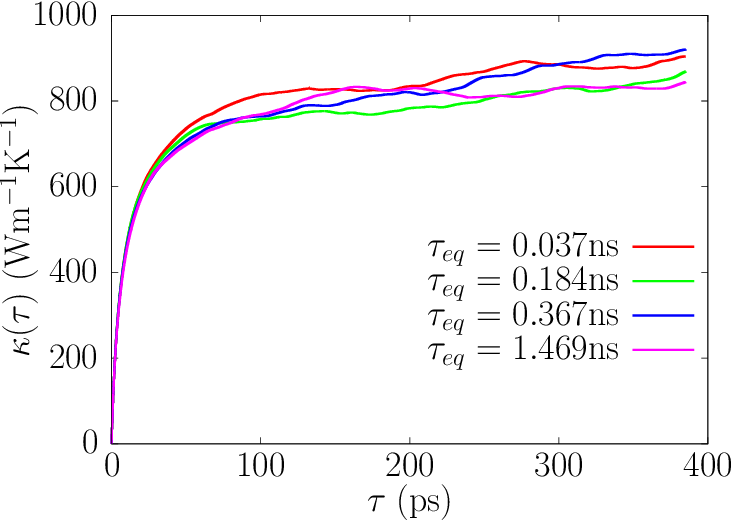} 
\caption{Thermal conductivity integral in Eq. \ref{GK} plotted as a function of the upper integration limit $\tau$ for different equilibration times $\tau_{eq}$. In each instance the system temperature was $T=300$K.
}
\label{fig1}
\end{centering}
\end{figure}

We begin with calculation of the standard Green-Kubo thermal conductivity given by the equation,
\begin{equation} \label{GK}
\kappa(\tau ) = {\Omega  \over 3 k_{B}T^{2}} \int_{0}^{\tau} \langle \bm{J}(t)\cdot \bm{J}(0) \rangle dt
\end{equation}
with the thermal conductivity formally given by taking the limit $\kappa =  \lim_{\tau \rightarrow \infty} \kappa(\tau)$.
In Fig. \ref{fig1}, the value of $\kappa(\tau)$ is shown as a function of the upper limit of integration $\tau$ for several different equilibration times.  The results in Fig. \ref{fig1} demonstrate well converged $\kappa$ values for $\tau_{eq}=0.184$ns and longer. Based on this, we use an equilibration time $\tau_{eq}=0.367$ns for each of the subsequent calculations. We determine $\kappa=(820 \pm 25)$Wm$^{-1}$K$^{-1}$ for this system size and $T=300$K for the longest equilibration time $\tau_{eq}=1.469$ns. This value for $\kappa$
appear to be very close to the room-temperature value $\kappa=849$Wm$^{-1}$K$^{-1}$ obtained for the same potential by Cai et al  \cite{Cai:2019aa}. 

\begin{figure}
\begin{centering}
\includegraphics[width=0.40\textwidth]{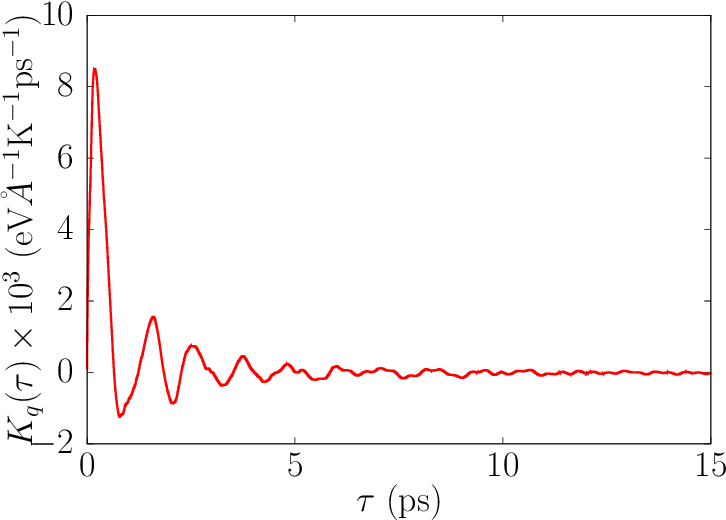} 
\caption{Response function $K_{\bm{q}}(\tau)$ plotted as a function of response time $\tau$ for $q={2 \pi \over \lambda}$ with $\lambda=6.93$nm. The system
temperature was $T=300$K. The oscillatory behavior is characteristic of second sound propagation.
}
\label{fig2}
\end{centering}
\end{figure}

For this very small system, the response function was computed for one vector $\bm{q}$ along the $\Gamma$-$M$ direction in the first Brillouin zone. The magnitude
of the $\bm{q}$ vector is given in terms of length scale $\lambda={2 \pi  \over q}$. 
Specifically, results are reported for  $\lambda=6.93$nm.  This can be compared to the value at the M-point at the zone edge of $\lambda_{M} = 0.43$nm. Hence, the vector $\bm{q}$ in each case lies closer to the $\Gamma$ point than the zone edge at $M$, and in fact describes thermal fluctuations periodic over a distance corresponding to $32$ primitive cells. 
The result for $K_{\bm{q}}(\tau)$ in Fig. 2 clearly shows wave-like transport characteristic of second sound. We conclude that for length scales $\sim 7$ nm and smaller at $T=300$K, transport is dominated by second sound propagation, with no evidence of diffusive transport. The decay in the response in Fig. \ref{fig2} is primarily due to a loss of phase coherence that arises with nonlinear dispersion and interference effects between different bands. Loss of coherence due to anharmonic scattering is another contribution, but Fig. \ref{fig2} shows a strong decay in second sound well before anharmonic scattering becomes important. This will be discussed in more detail after presenting the response function in Fourier space.

\begin{figure}
\begin{centering}
\includegraphics[width=0.40\textwidth]{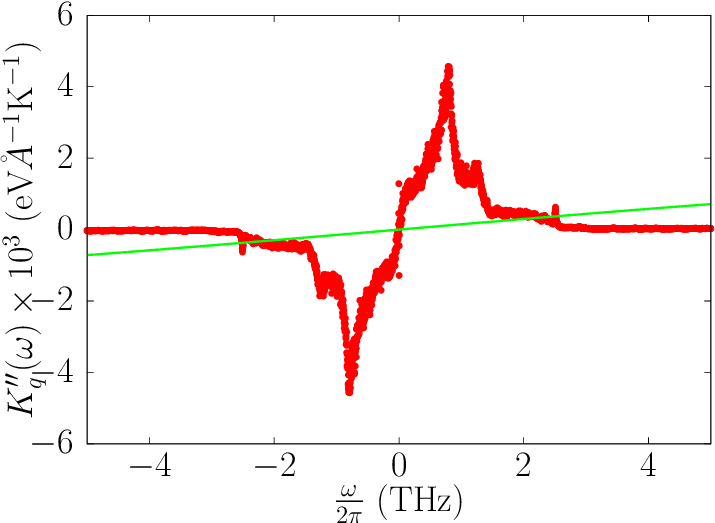} 
\caption{Response function $K^{\prime \prime}_{\bm{q}}(\omega)$ plotted as a function of frequency ${\omega \over 2 \pi}$. Results obtained for
$T=300$K and $q={2 \pi \over \lambda}$ with $\lambda=6.93$nm. The response based on the heat-diffusion equation is also shown (solid green line).
}
\label{fig3}
\end{centering}
\end{figure}

\begin{figure}
\begin{centering}
\includegraphics[width=0.40\textwidth]{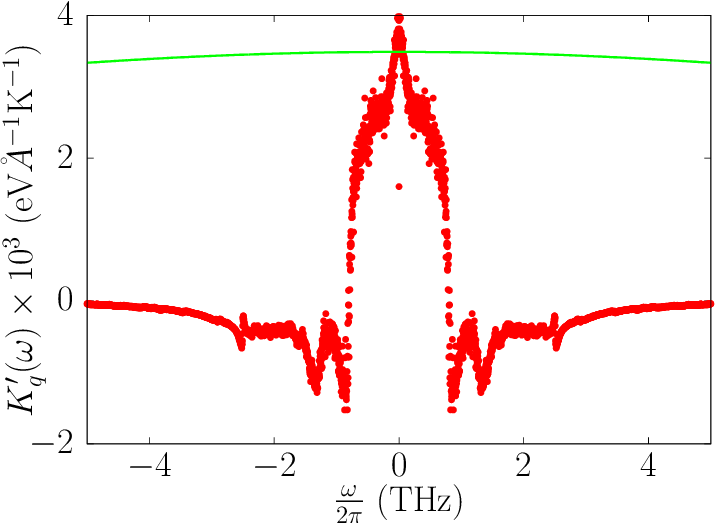} 
\caption{Real part of the response function $K^{\prime}_{\bm{q}}(\omega)$ plotted as a function of frequency ${\omega \over 2 \pi}$. Results obtained for
$T=300$K and $q={2 \pi \over \lambda}$ with $\lambda=6.93$nm. The response based on the heat-diffusion equation is also shown (solid green line).
}
\label{fig4}
\end{centering}
\end{figure}

\begin{figure}
\begin{centering}
\includegraphics[width=0.40\textwidth]{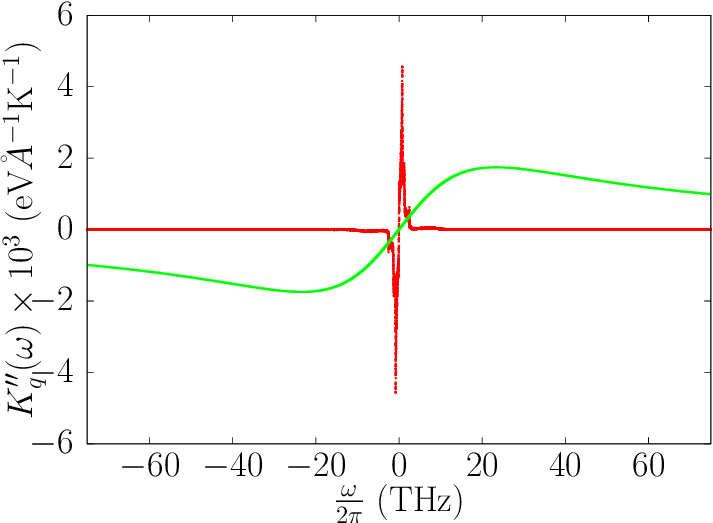} 
\caption{Response function $K^{\prime \prime}_{\bm{q}}(\omega)$ plotted as a function of frequency ${\omega \over 2 \pi}$. Results obtained for
$T=300$K and $q={2 \pi \over \lambda}$ with $\lambda=6.93$nm. The response based on the heat-diffusion equation is also shown (solid green line).
}
\label{fig5}
\end{centering}
\end{figure}

Further evidence for second sound propagation is obtained by Fourier transform of $K_{\bm{q}}(\tau)$ to obtain the frequency-dependent response function. The response function $K_{\bm{q}}(\tau)$ was determined over the range $0 \leq \tau \leq \tau_{m}$ with $\tau_{m}=385$ps. Well before the time $\tau = \tau_{m}$, the response function has decayed to zero. For this range of $\tau$ values, Fourier transforms were computed to determine the response function in frequency space using,
\begin{equation}
\tilde{K}_{\bm{q}}(\omega) = \int_{0}^{\tau_{m}} K_{\bm{q}}(\tau) e^{i \omega \tau}d\tau
\text{  ,} 
\end{equation}
Extending $\tau_{m}$ to larger values mainly would result in higher resolution, but would correspondingly require increased statistical averaging to minimize noise.
Whereas $ K_{\bm{q}}(\tau) $ is a real function, the Fourier transform results in real and imaginary components $K^{\prime}_{q}(\omega)$ and $K^{\prime \prime}_{\bm{q}}(\omega)$ respectively
so that $\tilde{K}_{\bm{q}}(\omega) = K^{\prime}_{q}(\omega)+i K^{\prime \prime}_{\bm{q}}(\omega)$. In Fig. \ref{fig3}, the imaginary component $K^{\prime \prime}_{\bm{q}}(\omega)$ is shown for $q={2\pi \over \lambda}$ with $\lambda=6.93$nm. Hence Fig. \ref{fig3} shows the imaginary part of  the Fourier transform of the data in Fig. \ref{fig2}. Also shown in Fig. \ref{fig3} is the 
response function obtained from an assumption that Fourier's Law applies (see Appendix B for a derivation). The results in Fig. \ref{fig3} demonstrate very strong deviations from the heat-diffusion equation for length scales $\lambda=6.93$nm at $T=300K$. The real component $K^{\prime}_{q}(\omega)$, which is related to $K^{\prime \prime}_{q}(\omega)$ via a Kramers-Kronig relation, is shown in Fig. \ref{fig4} also in comparison to the Fourier's law prediction. In Fig. \ref{fig5}, the same results are shown but with a wider range in frequency space to demonstrate the very strong disagreement with Fourier's law. 

The results in Fig. \ref{fig3} demonstrate that the decay in second sound observed in Fig. \ref{fig2} is primarily due to nonlinear dispersion and interference effects across the entire phonon spectrum rather than to anharmonic scattering events. First, assuming that there is a dominant peak in the response, the decay should be related to the width of the response function in frequency space. From Fig. \ref{fig3}, there is a dominant peak at about $\sim 1$THz, with a width $\sim 0.5$THz. Using the uncertainty relation, this would suggest lower limit for lifetime of second sound $\sim 0.5$ps. This is not inconsistent with the results in Fig. \ref{fig2}. Specifically, while clear oscillations persist to $\sim 5$ps, the signal is sharply attenuated after $1-2$ periods. The next question is what physics is responsible for the width in Fig. \ref{fig3}? First, the spectrum exhibits several rather sharp peaks which demonstrates the existence of persistent phase coherence. In Section 4 we will show that the details in the spectrum can be exactly identified by analysis of the phonon band structure. Hence, the width of the response function in Fig. \ref{fig3} is primarily described by the phonon band structure and only minimally by anharmonic scattering. We will see shortly that for longer length scales and higher temperatures, it becomes less clear what is the primary determinant for second-sound lifetime. However, nonlinear dispersion and interference between bands is always an important and possibly dominant effect.

 First, the detailed structure in Fig. \ref{fig3} demonstrates persistent phase coherence between normal modes. An analysis using the phonon band structure will be used later to elucidate the detailed structure in Fig. \ref{fig3}. If phonon modes scattered at a high enough rate for decoherence to occur, the structure seen in Fig. \ref{fig3} would eventually be lost. Finally, by applying the uncertainty relation to the response in Fig. \ref{fig3}, we would expect dominant oscillations at frequency $\sim 1$THz but with a lower limit to the lifetime of $\sim 0.5$ps. This is roughly consistent with the results displayed in Fig. \ref{fig2}. Hence we conclude that the sharp structure with identifiable peaks indicates persistent phase coherence between normal modes, and then the observed lifetime of second sound in Fig. \ref{fig2} is primarily due nonlinear dispersion and interference between different phonon bands. We note here that for longer length scales and higher temperature, as we will see shortly, often makes the relative roles of these different effects less clear. However, what we have shown is that the lifetime for second sound is not necessarily determine by anharmonic scattering, and in fact at short enough scales can be predominantly determined by interference between long-lived, phase coherent oscillations.

It is worthwhile to try to understand this very large disagreement, because it points to some fundamental aspects of how transport at short length scales deviates from Fourier's law and the heat-diffusion equation. Using the computed thermal diffusivity $\alpha=178.9$nm$^{2}$ps$^{-1}$, estimated from the computed $\kappa$ and the classical-equipartition theory for the heat capacity, and the wave number $q={2\pi \over \lambda}$ with $\lambda=6.93$nm, the relevant quantity in the heat-diffusion equation is ${\alpha q^{2} \over 2\pi} = 23.4$THz, which corresponds to a timescale $42.8$fs. This timescale is shorter than the time required for wave propagation at these length scales. A more complete analysis obtained from calculation of h-BN normal-mode dispersion relations is presented later. In short, the response function computed by MD simulations and shown in Figs. \ref{fig3}-\ref{fig5} can be understood entirely in terms of the normal-mode spectra, whereas the response obtained assuming the validity of Fourier's law predicts a featureless response over a very broad frequency range. 

Having established a baseline for calculations, larger system sizes and different temperatures were also explored. For transport corresponding to smaller values of $q$ and/or higher temperatures $T$, we expect to eventually observe scattering which destroys phase coherence and hence leads towards behavior eventually more consistent with Fourier's law. 
We specifically simulated a system with dimensions $512 \times 32$ primitive cells along the $\bm{a}_{1}$ and $\bm{a}_{2}$ Bravais lattice vectors with a total of $N=32,768$ atoms. This choice allows for studying fluctuations over relatively large length scales with reasonable computational cost. Here specifically we consider length scales $\lambda_{1}=110.8$nm, $\lambda_{2}=55.4$nm, and $\lambda_{3}=36.9$nm. These lengths correspond to wave vectors  with magnitudes $q_{1}=0.0567$nm$^{-1}$, $q_{2}=0.113$nm$^{-1}$, and $q_{3}=0.170$nm$^{-1}$. System temperatures $T=100$K, $T=300$K, and $T=1200$K were simulated. No change in the supercell dimensions were made for
higher temperatures despite the tendency for thermal contraction. Hence the calculated structures were all under mild tensile stress which gradually increases as $T$ increases. Due to the larger system size, while the equilibration time $\tau_{eq}=0.367$ns was still used, and  $64$ independent simulations were performed at each temperature, the total amount of data used for time averaging was reduced to $47.04$ns. Each independent calculation therefore includes $0.735$ns of simulation time which is smaller than what was used for the smaller system. As the data will demonstrate, this length of simulation time is enough for convergence of the response functions $K_{\bm{q}}(\tau)$. Specifically, $K_{\bm{q}}(\tau)$ tends to be quite small at $\tau \sim 50$ps, although for spectral resolution we extend integration times in Eq. 7 to a maximum time  $\tau_{m}=385$ps. The results demonstrate that $K_{\bm{q}}(\tau_{m})$ is quite small and hence each independent simulation is of sufficient length to resolve the response functions.

 \begin{figure}
\begin{centering}
\includegraphics[width=0.40\textwidth]{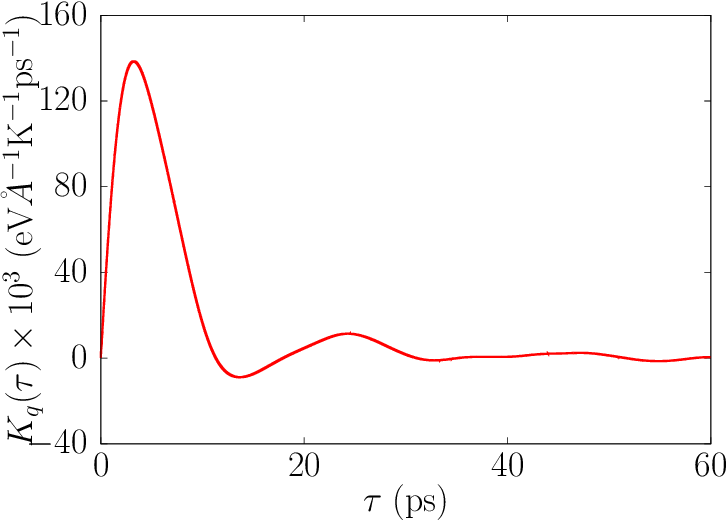} 
\caption{Response function $K_{\bm{q}}(\tau)$ plotted as a function of response time $\tau$ for $q=q_{1}={2 \pi \over \lambda_{1}}$ with $\lambda_{1}=110.8$nm. The system
temperature was $T=100$K.
}
\label{fig6}
\end{centering}
\end{figure}

\begin{figure}
\begin{centering}
\includegraphics[width=0.40\textwidth]{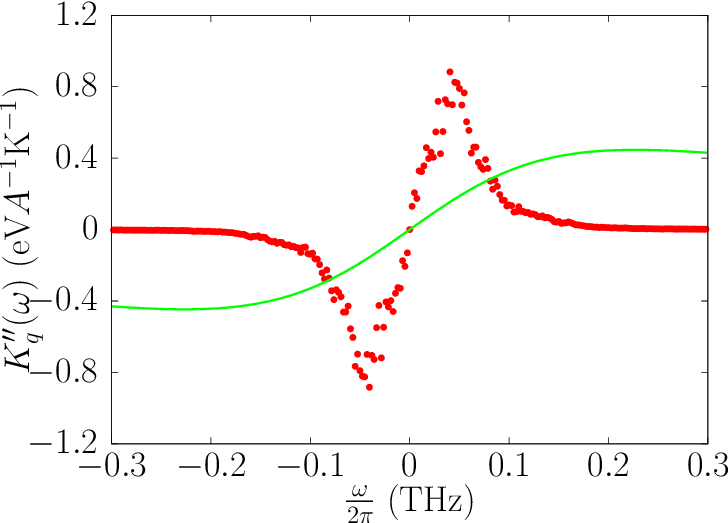} 
\caption{Imaginary part of the response function $K_{\bm{q}}^{\prime \prime} (\omega)$ plotted as a function of frequency ${\omega \over 2\pi}$ for $q=q_{1}={2 \pi \over \lambda_{1}}$ with $\lambda_{1}=110.8$nm (red dots). The system
temperature was $T=100$K. Comparison is made to predictions based on the heat-diffusion equation (solid green line).
}
\label{fig7}
\end{centering}
\end{figure}

At temperatures $T=100$K, second sound is observed for the longest length scale $\lambda_{1}=110.8$nm. Hence, here we focus on those results, recognizing that second sound will be present for any shorter length scale. In Fig \ref{fig6} the response function shows clear signatures of oscillation. In Fig. \ref{fig7}, $K_{\bm{q}}^{\prime \prime}(\omega)$ is plotted along with the response function predicted from the heat-diffusion equation. As with the shorter length scales, the failure the heat-diffusion equation to describe the data is evident.  From Fourier's law ${\alpha q_{1}^{2} \over 2\pi}=0.23$THz corresponding to a time scale of about $4.40$ps. By contrast, the MD result in Fig. \ref{fig7} show peaks at ${\omega \over 2\pi} \sim \pm 0.05$THz corresponding to a time scale of $\sim 20$ps which appears consistent with the dominant oscillation period in Fig. \ref{fig6}. Hence it is clear that the decay rate from Fourier's law is faster than wave propagation would allow. More detailed analysis follows, but the fact that wave propagation occurs on longer time scales is evident by the relatively narrow frequency range for $K_{\bm{q}}^{\prime \prime}(\omega)$ obtained by from the MD-simulated response function. Moreover,
it is clear that as length scale increases, the relevant time scale for the heat-diffusion equation rapidly increases and eventually surpasses the time required for wave propagation. 
Specifically, the time scale for diffusion increases quadratically with length scale, while the time scale for wave propagation increases linearly with length scale. Hence, the differences in Fig. \ref{fig7} for $\lambda_{1}=110.8$nm are much less dramatic than those seen in Fig. \ref{fig5} for the shorter length scale of $\lambda=6.93$nm.

 Another relevant observation from Fig. \ref{fig7} for $\lambda_{1}=110.8$nm is that fewer distinct resonant peaks are apparent in comparison to Figs. \ref{fig3} and \ref{fig5} for the shorter length scale. Not unexpectedly, wave propagation and second sound is more difficult to directly observe over longer distances even when transport is essentially ballistic. The less detailed structure in Fig. \ref{fig7} in part
 is likely due to increase decoherence at longer time scales, but also because the width of $\tilde{K}_{\bm{q}}(\omega)$ narrows with decreasing $q$ and the width of
 various peaks become larger than their separation. Possibly some detail could be resolved by using longer simulation times $\tau_{m}$. As with all other results,
 $\tau_{m}=385$ps was used for the Fourier transforms, and consequently the resolution is limited to about $2.4 \times 10^{-3}$THz or $2.4$ GHz. Any structure below this
 scale would require a larger time $\tau_{m}$ and probably better statistics to resolve. This is not pursued further  here.
 
\begin{figure}
\begin{centering}
\includegraphics[width=0.40\textwidth]{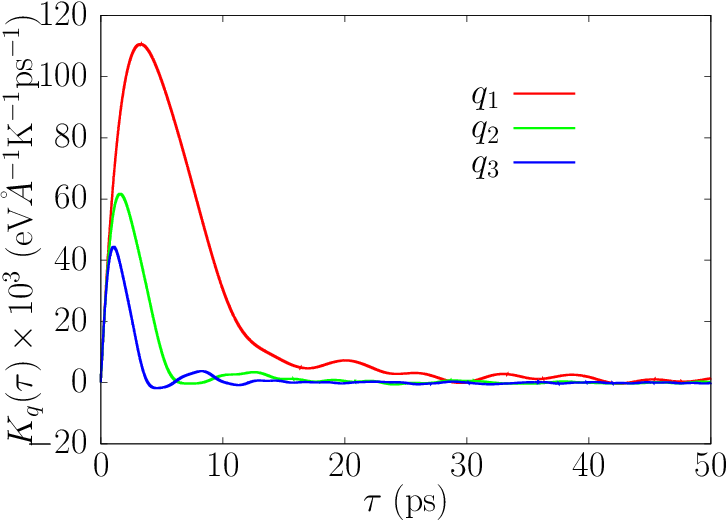} 
\caption{Response function $K_{\bm{q}}(\tau)$ plotted vs. $\tau$ for $q_{1}=0.0567$nm$^{-1}$, $q_{2}=0.113$nm$^{-1}$, and $q_{3}=0.170$nm$^{-1}$ for $T=300$K.
}
\label{fig8}
\end{centering}
\end{figure}

\begin{figure}
\begin{centering}
\includegraphics[width=0.40\textwidth]{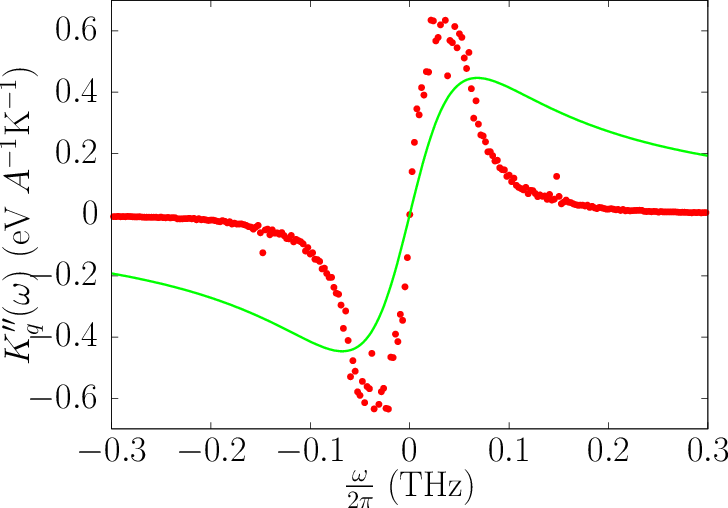} 
\caption{Response function $K^{\prime \prime}_{\bm{q}}(\omega)$ plotted vs. $\tau$ for $q=q_{1}=0.0567$nm$^{-1}$ for $T=300$K. Comparison is made to predictions based on the heat-diffusion equation (solid green line).
}
\label{fig9}
\end{centering}
\end{figure}

Given that second sound is observable for $\lambda_{1}=110.8$nm at $T=100$K, we next report results for three different scales at $T=300$K and $T=1200$K. The expectation is that anharmonic effects will reduce phase coherence and thereby eventually damp any oscillatory response. For $T=300$K, Fig. \ref{fig8} shows the response function as a function of time for each value $q_{1}$, $q_{2}$, and $q_{3}$. With the exception of $q_{3}$ which corresponds to the shortest length scale, second sound defined as displaying a change of sign in the response function is not observed. However, oscillatory behavior is evident in each case. The Fourier transform of the results for $\lambda_{1}={2 \pi \over q_{1}}= 110.8$nm is shown in Fig. \ref{fig9}. In addition to strong disagreement with the heat-diffusion equation in terms of the overall structure, the MD results show a distinct peak at ${\omega \over 2 \pi} = \pm 0.148$THz. This feature, which also is observed at $T=1200$K, is related to the LA branch as will be demonstrated and discussed in the next section. Another apparent difference in comparing the $T=100$K and $T=300$K results (Fig. \ref{fig7} and Fig. \ref{fig9}) is that the peak at the lower temperature $T=100K$ is somewhat sharper than the peak at $T=300$K. This indicates that, while nonlinear dispersion and interference between different bands is still an important consideration for the finite lifetime of second sound, at $T=300$K anharmonic scattering is beginning to be a more important effect.

\begin{figure}
\begin{centering}
\includegraphics[width=0.40\textwidth]{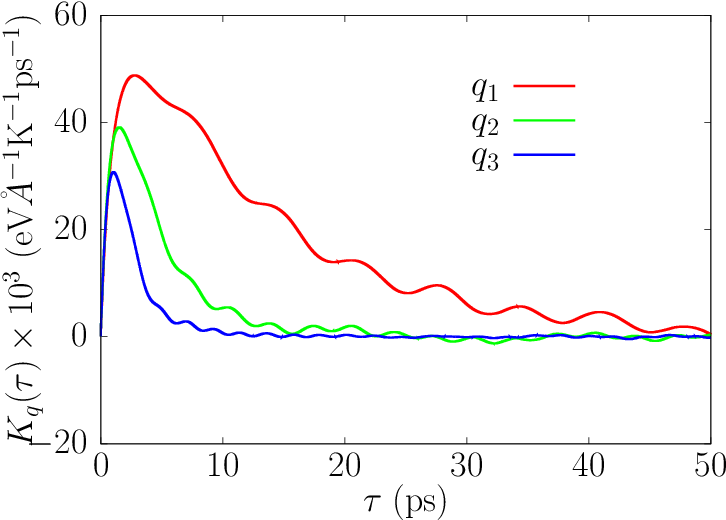} 
\caption{Response function $K_{\bm{q}}(\tau)$ plotted vs. $\tau$ for  $q_{1}=0.0567$nm$^{-1}$, $q_{2}=0.113$nm$^{-1}$, and $q_{3}=0.170$nm$^{-1}$  for $T=1200$K.
}
\label{fig10}
\end{centering}
\end{figure}

\begin{figure}
\begin{centering}
\includegraphics[width=0.40\textwidth]{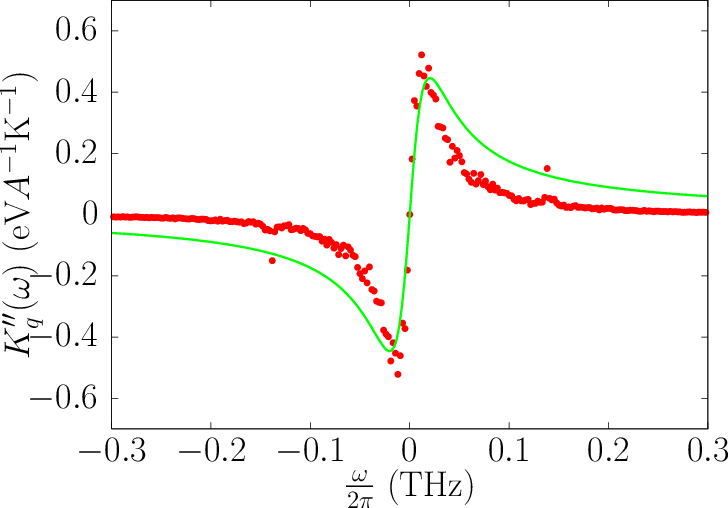}  
\caption{Response function $K^{\prime \prime}_{\bm{q}}(\omega)$ plotted vs. $\tau$ for $q=q_{1}=0.0567$nm$^{-1}$ for $T=1200$K. Comparison is made to predictions based on the heat-diffusion equation (solid green line).
}
\label{fig11}
\end{centering}
\end{figure}

For $T=1200$K simulations, oscillatory behavior is still apparent for $q_{1}$, $q_{2}$, and $q_{3}$ as the results in Fig. \ref{fig10} demonstrate. Fourier transforms demonstrate deviations from
the heat-diffusion equation even for $q_{1}={2 \pi \over \lambda_{1}}$ with $\lambda_{1}=110.8$nm. This is apparent from the data in Fig. \ref{fig11}. While the overall behavior at $T=1200$K is much closer to the heat-diffusion
equation, differences are apparent including small but very sharp resonant peaks at ${\omega \over 2\pi} \approx \pm 0.138$THz. This behavior is clearly is associated with the visible periodic oscillations
seen in Fig. \ref{fig10}. While this resonance is essentially second sound, due to increased anharmonicity, the response function in Fig. \ref{fig10} does not become negative, and second sound would be difficult to observed experimentally. Finally, apart from this oscillatory effect, the overall behavior is reasonably close to the heat-diffusion equation, despite evident differences. At $T=1200$K, completely diffusive heat transport would require a length scale greater than the largest scale explored here.

\section{Analysis}

We take the perspective that second sound specifically refers to an oscillatory response to a heat perturbation such that
the direction of the heat-flux density changes sign. In some of the results presented already, evidence of oscillation is present even without this feature, and in an experiment the oscillatory behavior would likely be difficult to observe. Ballistic transport is found
to some extent for all the simulations even at the largest length scale $\lambda_{1}=110.8$nm and highest temperature $T=1200$K. Here we attempt to identify specific contributions using analysis of the phonon band structure.

 To develop intuition, we first begin with the expression for the Fourier component  $\bm{J}_{\bm{q}}$ of the heat-flux density derived in Appendix C, and analogous to the expressions used in our previous work for one-dimensional chains \cite{Bohm:2022aa},
\begin{equation}
\bm{J}_{\bm{q}} = {1 \over 2\Omega} \sum_{\bm{k},s} \hbar \omega_{\bm{k}s} \bm{v}_{\bm{k}s} A_{\bm{k}+{\bm{q}\over 2},s}B_{-\bm{k}+{\bm{q}\over 2},s}
\text{  ,}
\end{equation}
in which $\omega_{\bm{k}s}$ is a phonon frequency for a mode with wave vector $\bm{k}$ in phonon branch $s$, with group velocity $\bm{v}_{\bm{k}s}$.
In a classical picture $A_{\bm{k}+{\bm{q}\over 2},s}$ and $B_{-\bm{k}+{\bm{q}\over 2},s}$ represent normal-mode coordinates. However, these might also represent
operators in a quantum-mechanical picture defined using phonon creation and annihilation operators. The crucial point to notice is that $\bm{J}_{\bm{q}}$ depends
on phase coherence between modes with wave vectors that differ by $\bm{q}$. For each pair of modes, we expect to observe
oscillatory behavior at a frequency given by the difference $\omega_{-\bm{k}+{1 \over 2} \bm{q},s}-\omega_{\bm{k}+{1 \over 2} \bm{q},s}$. In the case of
relatively small $\bm{q}$ in the region of band $s$ with nearly linear dispersion, we expect $\omega_{-\bm{k}+{1 \over 2} \bm{q},s}-\omega_{\bm{k}+{1 \over 2} \bm{q},s} \approx \omega_{\bm{q},s}$. This will lead to resonant behavior at frequency $\omega \approx \omega_{\bm{q},s}$ across a phonon band $s$ with linear dispersion. This behavior was
observed in the context of one-dimensional systems \cite{Bohm:2022aa} which show very strong resonance and persistent second sound.
 In two and three-dimensional crystals with more complex band structures, a much wider range of 
frequencies will be observed in contrast to a one-dimensional system. Given this wide range of frequencies, one would expect a corresponding shorter lifetime for second sound. In the computed response function $\tilde{K}_{\bm{q}}(\omega)$, this is the reason why there is clear structure but, with some exceptions noted, not necessarily sharp and isolated peaks. We note that some of these concepts have been discussed by others, including particularly Hardy\cite{Hardy:1963td} and also Sham in his early work on second sound\cite{Sham:1967aa,Sham:1967ab}, but
without clearly connecting the breadth of the spectrum and nonlinear dispersion to the lifetime of second sound.

\begin{figure}
\begin{centering}
\includegraphics[width=0.5\textwidth]{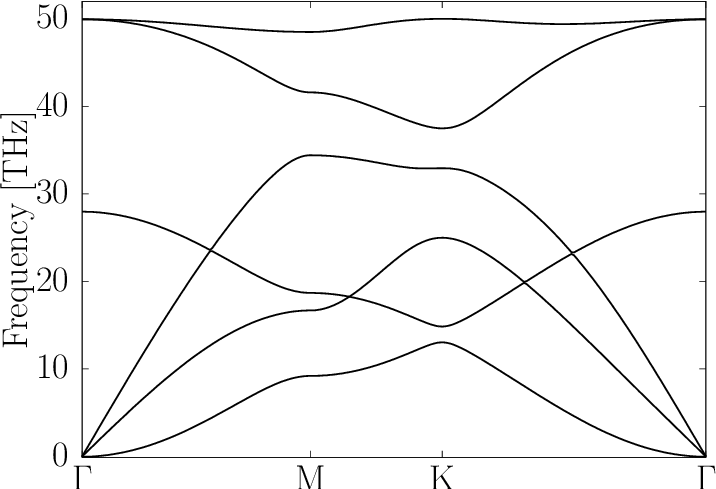} 
\caption{Computed phonon band structure for h-BN obtained from the empirical potential\cite{Lindsay:2011aa}.
}
\label{fig12}
\end{centering}
\end{figure}

The phonon band structure can be used to obtain insight into what modes contribute to different frequency ranges of the response function. 
We first compute the phonon band structure for h-BN using the empirical potential to describe interactions.  In Fig. \ref{fig12}, we show the computed phonon
band structure which was obtained using the phonon3py software package \cite{Togo:2015aa,Togo_2023}. The band structure is very similar to features in the bulk h-BN band structure from the original paper for the empirical potential used here \cite{Lindsay:2011aa}, and also shares common features with phonon band structures determined using first-principles DFT \cite{Li_2017}. Here we will focus only on modes with wave vectors directed along a $\Gamma \rightarrow M$ direction,
which corresponds to the chosen direction for $\bm{q}$ used in the MD simulations. Moreover, we focus in particular on the LA, TA, and ZA branches of the spectrum for 
relatively small wave vectors $\bm{k}$ along $\Gamma \rightarrow M$. For small wave vectors $\bm{k$}, the LA and TA branches exhibit nearly linear dispersion with slopes given by the sound LA and TA sound velocities, $v_{LA}=19.28$km s$^{-1}$ and $v_{TA}=10.98$km s$^{-1}$, respectively. The quadratic dependence on wave number $q$ for the ZA band yields group velocities in the range from zero to a maximum value of $v_{ZA}\approx 6.5$km s$^{-1}$. 

First we analyze the spectra in Fig. \ref{fig3} . Given the computed group velocities, a resonant peak associated with the LA branch should appear at $2.78$THz. Hence, the relatively sharp peak at about $2.71$THz must be related to the LA branch. For the TA branch, the predicted peak should occur at $1.58$THz, whereas in Fig. \ref{fig3} there appears to be a peak at around $1.35$THz. It is expected that, due to anharmonicity, resonant peaks undergo some frequency shifts from their predicted values based on $T=0$K phonon band structures. Finally, based on the computed range of ZA group velocities and the generally smaller optical phonon velocities, the spectra below $\sim 1$ THz should be comprised from those portions of the phonon spectra.

For the larger structure with $q_{1} ={2 \pi \over \lambda_{1}}$ and $\lambda_{1}=110.8$nm shown for $T=100$K in Fig. \ref{fig7}, only one very broad and prominent peak is evident at about $0.05$THz, and much of the detailed structure seen at shorter length scales is lost. This peak appears to be most consistent with the ZA branch of the spectra, although undoubtedly other contributions are present. Some contributions are evident in the range up to $\sim 0.20$THz, which is roughly consistent with the range where TA and LA branches should be evident. These results show that likely ZA modes are of predominant importance in second sound. However, given the quadratic dispersion of the ZA branch, the peak is very broad, and the second sound lifetime is rather short.

For the same structure and also  $q_{1} ={2 \pi \over \lambda_{1}}$ and $\lambda_{1}=110.8$nm , the spectra shown in Fig. \ref{fig9} for $T=300$K and Fig. \ref{fig11} for $T=1200$K, the same range of frequencies contribute to the response. However, as noted previously, these two cases also present evidence of a very sharp peak at $0.148$THz for $T=300$K, and $0.138$THz at $T=1200$K. The phonon spectra predicts the LA contribution at $0.174$THz. It is certain that these peaks represent the LA branch but with shifted frequencies by an amount that depends on temperature. We expect that these will be dependent on stress and thermal contraction. As noted elsewhere, the computed structures were not allowed to thermally contract, and hence these
systems are under slight tensile stress which increases with increasing $T$. However, the LA peaks themselves are small, and as with the $T=100$K spectra, the ZA modes represent a very large contribution.

Of course, LA and TA modes that have a nonzero component of $\bm{k}$ which is perpendicular to $\bm{q}$ will also contribute to the spectral response. However, this is not expected to lead to resonance behavior at a single frequency. For the ZA branch, since there is quadratic dispersion, coherent resonant response near a single frequency does not occur. 
Finally, optical modes also participate, primarily in the low-frequency range since the group velocities of optical modes are low compared to acoustic modes. Unlike LA and TA branches with $\bm{k}$ exactly directed along $\bm{q}$, these other contributions produce broad spectral response that is not resonant in the same way. These observations highlight the most significant factor in whether or not oscillatory heat transport can be observed experimentally.

It should finally be noted that while a very wide range of frequencies and nonlinear dispersion tend to result in short lifetimes for second sound, failure to observed second sound does not mean transport is necessarily accurately described by Fourier's law and the heat-diffusion equation. Rather, transport might still be ballistic, but with a lack of sufficient phase coherence across the spectrum to support second sound propagation. In this case, a description using the computed response functions will still be superior to assuming the validity of the heat-diffusion equation. Finally, we note that given these observations, it is evident that second sound should be more readily observed in relatively simple crystals primarily characterized by linear dispersion in the small $\bm{k}$ limit. Of course, low temperatures and short length scales are also relevant factors for the observation of second sound.

\section{Many-body perturbation theory and the Bethe-Salpeter equation}

The MD simulations reported above are tremendously useful for characterizing second sound and quantifying deviations from Fourier's law. However, there is a clear interest to develop a first principles, quantum-mechanical approach to this problem. The specific approach envisioned here begins with first-principles calculations based on density-functional theory (DFT) to determine produce phonon
band structures and lifetimes. Computing thermal response functions might then be accomplished using many-body perturbation theory. In fact, Sham has previously addressed this
problem resulting in an approach that depends on solutions of the Bethe-Salpeter equation \cite{Sham:1967aa,Sham:1967ab}. To the best of our knowledge, this prior theoretical work has not been implemented in calculations of realistic crystals. Here we present some aspects of this theory, which here helps to develop some physical intuition into thermal response functions computed from MD simulations. In Appendix C, a fairly extensive and complete derivation is provided.

In the proposed approach, the current-current correlation function is evaluated using Matsubara Green's functions. Following Sham\cite{Sham:1967aa,Sham:1967ab}, dressed phonon lines
are allowed to interact to obtain vertex corrections. This results in the Bethe-Salpeter equation which can be solved using the approximations developed by Sham \cite{Sham:1967aa,Sham:1967ab}. The resulting expression is based on iterative solution of the equation,
        \begin{equation} \label{zeq}
    Z_{K}(Q,\omega) =  {i W_{K\mu} +i \sum_{K^{\prime}} C_{K K^{\prime}} Z_{K^{\prime}}(Q,\omega)   \over  \omega - \bm{q}  \cdot \bm{v}_{K} + 2i \Gamma_{K} }
      \end{equation}
      in which $Z_{K}(Q,\omega)$ is the quantity to be solved. This object is clearly related to a dressed phonon Greens function. Also $Q$ represents $(\bm{q},s)$, and $K$, $K^{\prime}$ represent $(\bm{k},s)$, in which $s$ labels the phonon branch. The quantity $ C_{K K^{\prime}}$ is a scattering
      operator. Further details are provided in Appendix C including the definition for $W_{K\mu}$. 

The final expression for the imaginary component of response function is computed from
\begin{equation}
\begin{split}
K^{\prime \prime}_{\bm{q}} (\omega) & = 
{\beta \hbar \over 2 \Omega k_{B}T^{2} }
\sum_{K} {1 \over 2} \hbar \omega_{K} v_{K\mu} \left[n_{K}(n_{K}+1) \right]^{1 \over 2} \\
& \times  \left[ \left(n_{B}(\omega)+1 \right)\left(Z_{K}(Q,\omega)+Z^{*}_{K}(Q,\omega)  \right)+
n_{B}(\omega) \left(Z_{K}(Q,-\omega)+Z^{*}_{K}(Q,-\omega)  \right) \right] \text{   .}
\end{split}
\end{equation}
In this equation $n_{B}(\omega)$ represent the Bose-Einstein distribution function for a mode with frequency $\omega$.
The real part $K^{\prime}_{\bm{q}} (\omega)$ can then be obtained by using the Kramers-Kronig relation.

What is made physically clear by these equations is that a strong response is expected at frequency $\omega \approx \bm{q} \cdot \bm{v}_{K}$. This represents the frequency of coherent oscillations between two normal modes within the same phonon branch $s$ differing in wave vector by $\bm{q}$. 
One might refer to this as a ``beat'' phenomenon. Hence, this theory predicts
response functions in frequency space that can be directly connected to the phonon band structure as long as transport is in a ballistic regime. This point is directly consistent and verified by the MD simulations. Furthermore, in a band $s$ with nearly linear dispersion, such that the propagation speed is essentially constant ($v_{\bm{k}s} \approx v_{s}$ over a significant portion of reciprocal space), coherence can exist throughout the band. Hence, the finite lifetime of second sound is connected substantially to nonlinear dispersion and complexity of the phonon band structure, and not just to loss of coherence due to anharmonic scattering.

\section{Discussion and Conclusions}
In this paper we have presented MD simulation results for thermal transport in single-layer h-BN crystals at length scales up to $\sim 110$nm that demonstrate strong deviations from Fourier's law. At $T \sim 100$K for these length scales, second sound should be easily observable. At temperatures $T \sim 300$K and higher, second sound as defined by an oscillatory response may only be observable at length scales significantly below $\sim 110$nm. However, even at higher temperatures, transport tends to deviate quite strongly from Fourier's law and ballistic transport and wave transport is clearly evident. Specifically, the results demonstrate the role of finite wave propagation velocity. In fact, in the cases explored here, the characteristic time for heat diffusion is shorter than the wave propagation time. 
Comparison of the response functions computed from current-current correlation functions with those derived from Fourier's law and the heat-diffusion equation demonstrate the strong differences. Moreover, features in the response functions obtained from MD simulation are directly identifiable in relation to the normal-mode dispersion relations. In some cases, sharp resonant features are present even in cases where second sound likely would be difficult to observe.

The response function $\tilde{K}_{\bm{q}}(\omega)$ can show evidence of resonant phenomena. One key result obtained here is that width of resonant peaks is not only controlled by
finite phonon lifetimes and the associated broadening, but possibly more strongly by complexity in the phonon band structure and presence of non-linear dispersion. Moreover, as
$q$ decreases to smaller values, the separation between resonant peaks can often become smaller than the width of the peaks, thereby resulting in a simpler structure for $\tilde{K}_{\bm{q}}(\omega)$. This picture also points to the fact that it will be more likely to observe second sound in lower dimensional materials and also in materials characterized by linear phonon dispersions. 

Another relevant question is the role played by stress. As noted, the simulations corresponded to a small amount of tensile stress which increases with increasing temperature. It was noted in calculations of phonon lifetimes in graphene by Bonini et al \cite{Bonini_2012} that tensile stress can lead to a rapid increase in phonon lifetime. In fact, for systems without tensile stress, it was observed that lifetimes could be short enough in some cases to render some normal modes as poor quasiparticles for transport theory \cite{Bonini_2012}. The role of stress is certainly important in our work, and remains an area for future exploration. For example, the very sharp resonant peaks observed in calculations at $T=300$K and $T=1200$K and attributed to LA mode, shown in Fig. \ref{fig9} and Fig. \ref{fig11} respectively, may sharpen due to the increased tensile stress and a corresponding increase in the LA phonon lifetime.  It is noted that these peaks were not apparent at $T=100K$ in Fig. \ref{fig7} which could be due to the decreased tensile stress at that temperature. However, these points are conjectural and require further investigation to establish.

In current theories of second sound propagation, the governing equation is a damped wave equation for the evolution of the temperature field $T(\bm{r},t)$,
\begin{equation}
{\partial^{2} T \over \partial t^{2}}+ 
{1 \over \tau_{ss}} {\partial T \over \partial t} 
-v_{ss}^{2} \nabla^{2} T = 0 \text{ ,}
\end{equation}
typically referred to as the Cattaneo-Vernotte or hyperbolic heat equation. Theoretical analysis uses the PBE to determine
the propagation velocity $v_{ss}$ as an average over the acoustic propagation velocities and predictions for $\tau_{ss}$. Written
in terms of Fourier components of the temperature field $T_{\bm{q}}(t)$, the equation above become for each Fourier component,
\begin{equation}
{d^{2} T_{\bm{q}} \over dt^{2}}+
{1 \over \tau_{ss}} {d T_{\bm{q}} \over dt}
+ v_{ss}^{2} q^{2}T_{\bm{q}} = 0 \text{ ,}
\end{equation}
which is just the equation of a damped harmonic oscillator. Oscillations will be underdamped, and hence second sound observable,
for $q > {1 \over 2v_{ss}\tau_{ss}}$. However, this simple differential equation does not qualitatively describe the response functions obtained from the MD simulations reported here for reasons that are very clear. First, the hyperbolic heat equation in the underdamped regime predicts a response near frequency $\omega \sim v_{ss}q$ rather than the more complex structures observed here.  Second, due to the important role of phase coherence across the spectrum, the finite lifetime of oscillations is, to first order, controlled by nonlinear dispersion and the broad range of frequencies contributing the the phenomena. This important physical mechanism is not predicted from particle-based PBE solutions \cite{Hardy:1970aa,Lee:2017aa,Luo:2019aa,Shang:2022aa}. 
 
It is worth exploring the significance and meaning of the frequency dependence of the response function. First, the information captured by $\tilde{K}_{\bm{q}}(\omega)$ is related to the time-dependent response to a heat pulse with wave vector $\bm{q}$. As we have demonstrated, when transport is ballistic, it encodes propagation of phase-coherent phonon modes, with strong response at frequencies that correspond to underlying resonances. But another way to view $\tilde{K}_{\bm{q}}(\omega)$ is in terms of the heat-flux density that results from a spatially periodic, sinusoidal heating power $H_{\bm{q}}^{(ext)}(t)=\tilde{H}^{(ext)}_{\bm{q}}(\omega) e^{i \omega t}$. From Eq. \ref{def2}, the amplitude of the heat-current density in frequency space is,
\begin{equation}
\tilde{J}_{\bm{q}}(\omega) = -{i \over c_{V}}q \tilde{K}_{\bm{q}}(\omega)\tilde{H}^{(ext)}_{\bm{q}}(\omega)
\end{equation}
This expression makes it clear that the imaginary component ${K}^{\prime \prime}_{\bm{q}}(\omega)$ determines the amplitude of the heat-flux density which is in-phase with the heating source. Now one can give more thought to the practical meaning behind differences between the computed response functions and the predictions from Fourier's law. We restrict the discussion to ballistic transport regimes, either with or without second sound propagation. For source frequencies high enough to be outside of the range where coherent phonons have resonant frequencies, transport is less effective than predictions based on Fourier's law. Simply put, the phonons and their finite propagation velocity cannot keep up with high-frequency heat sources. For example, the results in Fig. \ref{fig6} for length scales $\lambda_{1}=110.8$nm show almost no response for frequencies greater than $\sim 0.15$THz. At longer length scales, for example $\lambda \sim 1\mu$m, the frequency range where effective transport occurs should decrease to $\sim 15$GHz, which is certainly relevant to many applications. By contrast, within the frequency range where resonant phonon transport occurs, results like those in Fig. \ref{fig6} indicate that
the response can be much stronger than what the heat-diffusion equation predicts. Finally, the phase relation between the heat-flux density and the heating source is substantially different from the expectations of the Fourier's law, something that we demonstrated in our original paper on thermal-response functions \cite{Fernando_2020}. In short, understanding the operation of devices in the ballistic regimes, response functions are needed to provide a clear picture of heat transport and dissipation.

Future efforts we expect to direct towards first-principles approaches based on the many-body perturbation theory outlined here. At low temperatures, where one expects second sound and ballistic transport to be
most relevant, quantum statistics is important. Currently we are working with the theoretical expressions here in practical calculations to address questions about convergence and numerics in general. These efforts will be the subject of future papers. 

The general outline of the theory provides important guidance on the basic physics of second sound and ballistic transport, and also what materials properties are most likely to correlated with the potential to observe second sound. First, as our paper on one-dimensional chains demonstrated, second sound can persist to very long times \cite{Bohm:2022aa}. This demonstrates that simple materials with linear dispersion relations are most likely to exhibit second sound. At low enough temperatures and for short enough scales, the lifetime for second sound is limited primarily by the presence of nonlinear dispersion and complex band structures, both of which lead to a rapid loss of phase coherence. 
As a result of these facts, in contrast to one-dimensional chains, two and three-dimensional materials are less likely to support long-lived second sound. For a first approximation, one can apply the uncertainty principle to the peak structure in the response function in frequency space to estimate the lifetime for second sound.

For longer length scales and higher temperatures, the peaks in the response function in frequency space tends to become broader with fewer resolved features. As length scales increase, corresponding to smaller values of the wave vector $\bm{q}$, the spacing between sharp features in $\tilde{K}_{\bm{q}}(\omega)$ diminishes. In addition, as the magnitude of $\bm{q}$ decreases, the characteristic period of second sound increases. Higher temperatures increase the anharmonic scattering rate which reduces the lifetime of the normal modes, resulting in phase decoherence. This results in broader features in the response function $\tilde{K}_{\bm{q}}(\omega)$. For example, this is apparent from a comparison between Fig. \ref{fig7} and Fig. \ref{fig9} for $T=100$K and $T=300$K respectively. Applying the uncertainty principle, broader, less well-defined peak structures correspond to reduced lifetime of second sound. When the characteristic frequencies for second sound become comparable to the peak width in $\tilde{K}_{\bm{q}}(\omega)$, the lifetime for second sound becomes comparable to the period. It is the point where the lifetime and the period of second sound become comparable that Fourier's law and the heat-diffusion equation become a reasonably good description of heat transport.

In short, as $\bm{q}$ decreases in magnitude, the period of second sound increases. For lower values of $\bm{q}$ then, as the period for second sound becomes longer, loss of phase coherence, which is observed as broader peaks, is correlated with dissipation of second sound. In other words, longer periods for the second sound require the presence of sharper peaks in $\tilde{K}_{\bm{q}}(\omega)$ for second sound to persist. While smaller $\bm{q}$ values tends compresses  sharp features in $\tilde{K}_{\bm{q}}(\omega)$, this can occur in such a way that sharp features are lost. Moreover, as temperature increases, anharmonic scattering leads to decoherence and tends to broaden features in $\tilde{K}_{\bm{q}}(\omega)$. These two observations of the trend with smaller $\bm{q}$ values and higher T explains how second sound is difficult to observe over long length scales and at higher T. Another equivalent way to view this is the often-used heuristic that a minimal requirement for ballistic transport and second sound is that the mean free path for phonons must be longer than the length scale $\lambda = {2\pi \over q}$. Either increasing the length scale (by decreasing $\bm{q}$) or increasing the scattering rate (by increasing $T$) will tend to lead to dissipation of second sound.

However, even when
second sound has too short a lifetime to be observed, response functions remain a way to compute transport in ballistic regimes where the heat-diffusion equation does not apply. Finally, while the
method here specifically applies to infinite periodic systems, there are ways to incorporate point defects into Green's function approaches, and moreover it should be possible to address both superlattice structures and different kinds of boundary conditions. Those remain additional areas to be explored.

\section{Acknowledgements} 
PKS would like to acknowledge Prof. Eduardo Mucciolo of the University of Central Florida (UCF) for very useful discussions about the many-body perturbation theory approach including solutions
of the Bethe-Salpeter equation. We also acknowledge support and resources provided by UCF and the STOKES computer run by the UCF-ARCC administered by the Institute of Simulation and Training. All of the simulation data was obtained using this facility.

\newpage

\section*{Appendix A}
 \renewcommand{\theequation}{A\arabic{equation}}
  \setcounter{equation}{0}

Here we derive expressions for the local energy density, heat-current density, including representations in Fourier space. We start with the excess thermal energy of an atom $i$
\begin{equation}
\epsilon_{i} = E_{i}-E_{0} \text{   ,}
\end{equation}
in which $E_{i}$ is the energy of atom $i$ and $E_{0}$ is the $T=0$K bond energy of an atom. Defined in this way, $\epsilon_{i}$ represents only the local thermal energy. Then we can define energy-density
expressions,
\begin{equation}
\epsilon(\bm{r}) = \sum_{il}   \epsilon_{i}\delta^{(3)}  (\bm{r} - \bm{r}^{(0)}_{i} ) =\sum_{\bm{q}} \epsilon_{\bm{q}} e^{i \bm{q} \cdot \bm{r}}  \text{   .}
\end{equation}
Then the energy density is given in Fourier space as,
\begin{equation}
\epsilon_{\bm{q}}= {1 \over \Omega}\sum_{i} \epsilon_{i}e^{-i \bm{q} \cdot   \bm{r}^{(0)}_{i}} \text{   .}
\end{equation}
For the form of $E_{i}$, we use the expressions presented in Ref.\cite{Fan:2015tj},
\begin{displaymath} \label{locen}
E_{i} = {1 \over 2} m_{i}\bm{v}_{i} \cdot \bm{v}_{i} + 
{1 \over 2} \sum_{j \ne i} U_{ij} \text{  ,}
\end{displaymath}
in which the velocities are given by
 $\bm{v}_{i} =\dot{\bm{u}}_{i}$ and the form for the potential energy is taken from the Tersoff potential
 \cite{Tersoff:1989tr,Fan:2015tj},
 \begin{equation}
 U_{ij}=f_{c}(r_{ij}) \left[f_{R}(r_{ij}) - b_{ij} f_{A}(r_{ij})\right]  \text{  .}
 \end{equation}
 In this expression $r_{ij}$ is the distance between atoms $i$ and $j$ and $b_{ij}$ is a bond-order
 term which depends on the local environment \cite{Tersoff:1989tr}. The complete form for the Tersoff potential 
 is given in many other articles including Ref. \cite{Tersoff:1989tr}. The total energy of the system is given by,
 \begin{equation}
 E_{tot} = \sum_{i} E_{i}  \text{   .}
  \end{equation}
  
  Next the heat-flux density can be represented as,
 \begin{equation}
\bm{J}(\bm{r}) = \Omega \sum_{i} \bm{J}_{i} \delta^{(3)} (\bm{r}-\bm{r}^{(0)}_{i})=
\sum_{\bm{q}} \bm{J}_{\bm{q}} e^{i \bm{q} \cdot \bm{r}}  \text{   ,}
\end{equation}
and the Fourier-space heat-flux density is,
 \begin{equation}
\bm{J}_{\bm{q}} = \sum_{i} \bm{J}_{i} e^{- i \bm{q} \cdot \bm{r}^{(0)}_{i}}  \text{  .} 
\end{equation}
In these definitions, we use the $T=0$K atom coordinates $\bm{r}_{i}^{(0)}$. Using these coordinates
in the definitions here results in no contributions due to the so-called ``convective'' terms.

Applying the continuity equation in Fourier space, we obtain the equation,
 \begin{equation} \label{conteq}
{d \epsilon_{\bm{q}} \over dt} = -i\bm{q} \cdot \bm{J}_{\bm{q}} \text{   .}
\end{equation}
This makes it clear that the projection of $\bm{J}_{\bm{q}}$ along the direction of $\bm{q}$ is relevant, and for simplicity we write,
 \begin{equation} \label{conteq2}
{d \epsilon_{\bm{q}} \over dt} = -iq J_{q} \text{   ,}
\end{equation}
in which $q=|\bm{q}|$ and $J_{q}$ is defined by,
 \begin{equation}
J_{q} = {\bm{q} \cdot \bm{J}_{\bm{q}} \over q}   \text{  .} 
\end{equation}

To determine the expression for the local heat current associated with each site, we start by evaluating
the time derivative of $\epsilon_{\bm{q}}$,
 \begin{equation} \label{dlocen}
{d \epsilon_{\bm{q}} \over dt} =
{1 \over \Omega} \sum_{i} {dE_{i} \over dt} e^{-i \bm{q} \cdot  \bm{r}^{(0)}_{i} } 
\end{equation}
The time derivative of the local energy $E_{i}$ is,
 \begin{equation}
{dE_{i} \over dt} = \bm{F}_{i} \cdot \bm{v}_{i} + {1 \over 2} \sum_{j \ne i} {dU_{ij} \over dt} \text{   ,}
\end{equation}
in which $\bm{F}_{i}$ is the total force for atom $i$. This is then input into Eq. \ref{dlocen} and the terms rearranged,
 \begin{equation} 
 \begin{split}
{d \epsilon_{\bm{q}} \over dt} = &
- {1 \over 2 \Omega} \sum_{j \ne i} \sum_{\mu} \left[{\partial U_{ij} \over \partial r_{i\mu}}v_{i \mu} e^{-i \bm{q} \cdot \bm{r}^{(0)}_{i}} \right]
 \left[1-e^{i \bm{q} \cdot(\bm{r}^{(0)}_{i} - \bm{r}^{(0)}_{j} )} \right] - \\
 &{1 \over 2 \Omega} \sum_{j \ne i} \sum_{k \ne j,i} \sum_{\mu} \left[{\partial U_{jk} \over \partial r_{i\mu}}v_{i \mu} e^{-i \bm{q} \cdot \bm{r}^{(0)}_{i}} \right]
 \left[1-e^{i \bm{q} \cdot(\bm{r}^{(0)}_{i} - \bm{r}^{(0)}_{k})} \right]
 \text{   .}
 \end{split}
\end{equation}
The assumption used next is that $\bm{q}$ is small such that the terms in brackets can be expanded to lowest order. This yields,
 \begin{equation} 
 \begin{split}
{d \epsilon_{\bm{q}} \over dt} = & 
-{1 \over 2 \Omega} \sum_{j \ne i} \sum_{\mu} \left[i \bm{q} \cdot \bm{r}^{(0)}_{ji}{\partial U_{ij} \over \partial r_{i\mu}}v_{i \mu} e^{-i \bm{q} \cdot \bm{r}^{(0)}_{i}} \right]
- \\
 &{1 \over 2 \Omega} \sum_{j \ne i} \sum_{k \ne j,i} \sum_{\mu} \left[i \bm{q} \cdot \bm{r}^{(0)}_{ki}{\partial U_{jk} \over \partial r_{i\mu}}v_{i \mu} e^{-i \bm{q} \cdot \bm{r}^{(0)}_{i}} \right]
 \text{   ,}
 \end{split}
\end{equation}
in which $\bm{r}^{(0)}_{ij}=\bm{r}^{(0)}_{i}-\bm{r}^{(0)}_{j}$.
Comparison to Eq. \ref{conteq}, it is clear that the local heat-current associated with atom $i$ is,
 \begin{equation}
\bm{J}_{i} =  
-{1 \over 2\Omega} \sum_{j \ne i} \bm{r}^{(0)}_{ij}  
{\partial U_{ij} \over \partial \bm{r}_{i}} \cdot \bm{v}_{i} 
 -{1 \over 2\Omega} \sum_{j \ne i} \sum_{k \ne i,j}\bm{r}^{(0)}_{ik}
{\partial U_{jk} \over \partial \bm{r}_{i}} \cdot \bm{v}_{i}   \text{  .} 
\end{equation}

\section*{Appendix B}
 \renewcommand{\theequation}{B\arabic{equation}}
  \setcounter{equation}{0}
  
Here we derive the relevant expressions for the response functions $K^{\prime}_{\bm{q}}(\omega)$ and $K^{\prime \prime}_{\bm{q}}(\omega)$ that correspond to Fourier's law and the heat-diffusion equation. We also relate these expressions to the standard Green-Kubo calculation of thermal conductivity $\kappa$.

In the case of Fourier's law, it is easily shown that the response function is given by $K_{\bm{q}}(t) = \kappa e^{-\alpha q^{2}t}$. The thermal diffusivity is $\alpha={\kappa \over c_{V}}$ . However, in Fourier's
law, the Fourier component of the heat-flux density $J_{\bm{q}}(t)$ is directly proportional to the energy perturbation $u_{\bm{q}}(t)$ at the same time $t$ independent of the history. Hence,
assuming there is only one heat pulse input $u_{\bm{q}}(0)$ at time $t=0$, the current at subsequent times is,
\begin{equation}
J_{\bm{q}}(t) = -{i \over c_{V}} q K_{\bm{q}}(t) u_{\bm{q}}(0) = -i \alpha q u_{\bm{q}}(t)
\end{equation}
Multiplication by $u_{-\bm{q}}(0)$ and taking an ensemble average, this results in the expression,
\begin{equation}
K_{\bm{q}}(t) = \kappa { \langle u_{\bm{q}}(t) u_{- \bm{q}}(0)  \rangle \over \langle u_{\bm{q}}(0) u_{- \bm{q}}(0)  \rangle }
\end{equation}
Using the known expression for $K_{\bm{q}}(t)$ from solving the heat diffusion equation this is just,
\begin{equation}
e^{-\alpha q^{2}t} =   { \langle u_{\bm{q}}(t) u_{- \bm{q}}(0)  \rangle \over \langle u_{\bm{q}}(0) u_{- \bm{q}}(0)  \rangle }
\end{equation}
Taking a second derivative with respect to time of both sides and using the continuity equation, we obtain,
\begin{equation}
\alpha  {d \over dt} e^{-\alpha q^{2}t}  = - { \langle J_{\bm{q}}(t) J_{- \bm{q}}(0)  \rangle \over \langle u_{\bm{q}}(0) u_{- \bm{q}}(0)  \rangle }
\end{equation}
Integration of this expression yields the expression, after taking the limit $\bm{q} \rightarrow 0$, the expression for thermal conductivity
in Eq. \ref{GK}.
The thermal fluctuations in the $\bm{q} \rightarrow 0$ limit are related to the volumetric heat capacity $c_{V}$ via,
\begin{equation}
{1 \over c_{V}} \lim_{\bm{q} \rightarrow 0} \langle u_{\bm{q}}(0) u_{- \bm{q}}(0)  \rangle = {k_{B}T^{2} \over V}
\end{equation}

Finally, the quantities $K_{\bm{q}}^{\prime}(\omega)$ and $K_{\bm{q}}^{\prime \prime}(\omega)$ corresponding to the real and imaginary 
components of the response function in frequency space are given. These are obtained from Fourier transform of $K_{\bm{q}}(t)=\kappa e^{-\alpha q^{2}t}$, and are given by
\begin{equation}
K_{\bm{q}}^{\prime}(\omega) = \kappa{\alpha q^{2} \over (\alpha q^{2})^{2} + \omega^{2}}
\end{equation}
\begin{equation}
K_{\bm{q}}^{\prime \prime}(\omega)=\kappa{\omega \over (\alpha q^{2})^{2} + \omega^{2}}
\end{equation}

\section*{Appendix C}
 \renewcommand{\theequation}{C\arabic{equation}}
  \setcounter{equation}{0}
  
The definition of the Fourier space heat current for many-body perturbation theory calculations also requires a definition of a local energy for each atomic site. We 
define the harmonic Hamiltonian in terms of momenta $\bm{p}_{il}$ and displacements from equilibrium $\bm{u}_{il}$ as,
 \begin{equation}
 H_{0} = { p_{il}^{2} \over 2m_{i}} + {1 \over 2} \sum_{i\alpha,j \beta} \Phi^{(2)}_{il\alpha,jl^{\prime}\beta} u_{il \alpha} u_{jl^{\prime} \beta}   \text{   ,}
\end{equation}
in which  $m_{i}$ represents the mass of atom $i$, and $u_{i}$ and $\Phi^{(2)}_{il\alpha,jl^{\prime}\beta}$ is the force-constant matrix. The indices $l$ and $l^{\prime}$ label primitive unit cells, and $\alpha$ and $\beta$ label Cartesian components. Following Hardy \cite{Hardy:1963td} and
others, the local site energy is defined as,
 \begin{equation}
 E_{il}={ p_{il}^{2} \over 2m_{i}} + {1 \over 2} \sum_{j^{\prime}} \sum_{\alpha, \beta} \Phi^{(2)}_{il\alpha,jl^{\prime}\beta} u_{il \alpha} u_{jl^{\prime} \beta}   \text{   .}
\end{equation}
Following a derivation analogous to the one given in Appendix A, and focussing only on the harmonic terms, the local current is given by,
\begin{equation}
\bm{J}_{il} = -{1 \over 2m_{i} \Omega} \sum_{j l^{\prime}}\sum_{\alpha, \beta} \left(\bm{x}_{ij}+\bm{R}_{ll^{\prime}} \right) p_{i l\alpha}  \Phi^{(2)}_{i\alpha,j\beta}  u_{jl^{\prime} \beta}   \text{   .}
\end{equation}
In this expression, $\bm{R}_{ll^{\prime}} =\bm{R}_{l}-\bm{R}_{l^{\prime}}$, in which $ \bm{R}_{l}$ and $\bm{R}_{l^{\prime}}$ are Bravais lattice vectors.
Then the Fourier space current is given by,
\begin{equation} \label{jq1}
\bm{J}_{\bm{q}} = \sum_{il}\bm{J}_{il} e^{-i \bm{q} \cdot ( \bm{x}_{i}+\bm{R}_{l})} =
 -{1 \over 2\Omega}  \sum_{il} {1 \over m_{i}} e^{-i \bm{q} \cdot ( \bm{x}_{i}+\bm{R}_{l})} \sum_{jl^{\prime}} \sum_{\alpha, \beta}\left(\bm{x}_{ij}+\bm{R}_{ll^{\prime}} \right) p_{i l\alpha} \Phi^{(2)}_{il\alpha,jl^{\prime}\beta}  u_{jl^{\prime} \beta}  \text{   .}
\end{equation}
We define the dynamical matrix for wave vector $\bm{k}$ as,
\begin{equation}
D_{i\alpha, j \beta}(\bm{k}) = \sum_{l^{\prime}} {\Phi^{(2)}_{il\alpha,jl^{\prime}\beta}   \over (m_{i}m_{j})^{1 \over 2}} e^{i \bm{k} \cdot(\bm{x}_{ji} +\bm{R}_{l^{\prime}l})}
\end{equation}
Next the displacement and momentum vectors are represented in a basis of harmonic eigenstates of the dynamical matrix,
\begin{equation}
u_{jl^{\prime}\beta} = \sum_{\bm{k},s}\left({\hbar \over 2m_{i} \omega_{\bm{k}s}} \right)^{1 \over 2}  A_{\bm{k}s} 
\epsilon_{\bm{k}s,j \beta} e^{i \bm{k} \cdot(\bm{x}_{j} + \bm{R}_{l^{\prime}})}
\end{equation}
and,
\begin{equation}
p_{il\alpha} =-i\sum_{\bm{k},s}\left({m_{i}\hbar \omega_{\bm{k}-\bm{q}s} \over 2  } \right)^{1 \over 2}  B_{-\bm{k}+\bm{q}s} 
\epsilon_{-\bm{k}+\bm{q}s,i \alpha} e^{-i (\bm{k}-\bm{q}) \cdot(\bm{x}_{i} + \bm{R}_{l})}
\end{equation}
in which $s$ is a phonon band index. 
Substitution of these expressions into Eq. \ref{jq1} and then noting, as in Ref. \cite{Hardy:1963td}, we obtain the relation,
\begin{equation}
2 \omega_{\bm{k}s} {\partial \omega_{\bm{k}s} \over \partial k_{\gamma}} = 2 \omega_{\bm{k}s}  v_{\bm{k} s \gamma}
\approx -i \sum_{il} \sum_{jl^{\prime}} \sum_{\alpha,\beta} (x_{ij \gamma} + R_{ll^{\prime} \gamma}) 
\epsilon_{\bm{k}s,j \beta} {\Phi^{(2)}_{il\alpha,jl^{\prime}\beta}   \over (m_{i}m_{j})^{1 \over 2}} \epsilon_{-\bm{k}+\bm{q}s,i \alpha} 
e^{i \bm{k} \cdot(\bm{x}_{ji} +\bm{R}_{l^{\prime}l})} \text{  .} 
\end{equation}
Then, using the fact that $\bm{q}$ is a small quantity, we obtain the final result,
\begin{equation} \label{curr}
\bm{J}_{\bm{q}} = {1 \over 2\Omega} \sum_{\bm{k},s} \hbar \omega_{\bm{k}s} \bm{v}_{\bm{k}s} A_{\bm{k}+{\bm{q}\over 2},s}B_{-\bm{k}+{\bm{q}\over 2},s}
\text{  ,}
\end{equation}
which was also given by Hardy \cite{Hardy:1963td}.
In writing this expression, the assumption that $\bm{q}$ is a small quantity leads to the reasonable approximation $\omega_{\bm{k}-\bm{q}s} \approx \omega_{\bm{k}s}$. The phonon operators are written in terms of the creation and annihilation operators,
\begin{equation} 
A_{\bm{k}s}=a_{\bm{k}s}+a^{\dagger}_{-\bm{k}s} \text{  ,}
\end{equation}
and,
\begin{equation} 
B_{\bm{k}s}= a_{-\bm{k}s}^{\dagger} - a_{\bm{k}s} \text{  .}
\end{equation}
It should be noted that the above neglects the so-called non-diagonal contributions \cite{Hardy:1963td} which mix different phonon branches together. The neglected terms, 
with the exception of instances where bands cross or nearly cross, are assumed to yield rapidly oscillating terms to the heat-flux density which contribute little 
to conduction.
Finally, as with the expressions derived in Appendix A, the relevant heat-flux density is given by the projection of $\bm{J}_{\bm{q}}$ along $\bm{q}/q$.

To address scattering due to anharmonicity, we assume the Hamiltonian can be written with harmonic and cubic anharmonic terms written using the phonon operators $A_{\bm{k}s}=a_{\bm{k}s}+a^{\dagger}_{-\bm{k}s}$,
\begin{equation} \label{hamiltonian}
H ={1 \over 2} \sum_{\bm{k},j} \hbar \omega_{\bm{k}s} A_{\bm{k}s} A_{-\bm{k}s}
+ \sum_{\bm{k}_{1},\bm{k}_{2},\bm{k}_{3}}\sum_{s_{1},s_{2},s_{3}}V^{(3)}(\bm{k}_{1}s_{1},  \bm{k}_{2}s_{2},\bm{k}_{3}s_{3})
 A_{\bm{k}_{1}s_{1}}  A_{\bm{k}_{2}s_{2}}  A_{\bm{k}_{3}s_{3}} \text{  .}
\end{equation}
Here the $a_{\bm{k}s}$ and $a^{\dagger}_{\bm{k}s}$ represent phonon annihilation and creation operators, respectively, for phonons with wave vector $\bm{k}$ and band labelled by $s$. In the following, the harmonic part of the Hamiltonian is represented by $H_{0}$, and the cubic anharmonic part is represented by $H_{A}$.

We next define a function using the Matsubara approach,
\begin{equation} \label{feq1}
\mathscr{C}_{\bm{q}}(\tau) ={1 \over 2} \langle T_{\tau}
J_{\bm{q}}(\tau) J_{-\bm{q}}(0)  \rangle   \text{  ,}
\end{equation}
with the assumption that $\bm{q}=q \hat{e}_{\mu}$ with the Cartesian direction $\hat{e}_{\mu}$ directed along a principle axis of the crystal.
 Here $T_{\tau}$ is an ordering operator, and $\tau$ is a real variable defined by treating time as a complex temperature.  The connection between Eq. \ref{feq1} and the thermal response function is described in detail in Appendix A.
The current-current correlation function is evaluation using the perturbation series expansion,
\begin{equation} \label{feq2}
\mathscr{C}_{\bm{q}}(\tau) ={1 \over 2} \langle T_{\tau}
\tilde{J}_{\bm{q}}(\tau) \tilde{J}_{-\bm{q}}(0) \sum_{n=0}^{\infty} {(-1)^{n} \over n!} 
 \int_{0}^{\beta} d\tau_{1} ... \int_{0}^{\beta} d\tau_{n} \tilde{H}_{A}(\tau_{1}) ... \tilde{H}_{A}(\tau_{n}) \rangle_{0}   \text{  ,}
\end{equation}
in which the operators evolve in the interaction representation according to,
\begin{equation}
\tilde{O}(\tau) = e^{\tau H_{0}} O e^{-\tau H_{0}}   \text{  .}
\end{equation}
The operator $H_{0}$ here is just the harmonic part from Eq. \ref{hamiltonian}. Write this result using the current-density operator defined in Eq. \ref{curr},
\begin{equation} \label{feq3}
\begin{split}
\mathscr{C}_{\bm{q}}(\tau)  & ={1 \over 4\Omega^{2}} \sum_{KK^{\prime}}{1 \over 2} \hbar \omega_{K} v_{K \mu}  \hbar \omega_{K^{\prime}}
v_{K^{\prime} \mu}   \langle T_{\tau} \tilde{A}_{K_{1}}(\tau) \tilde{A}_{\bar{K}^{\prime}_{1}}(0) \tilde{B}_{K_{2}}(\tau) 
\tilde{B}_{\bar{K}^{\prime}_{2}}(0)  \\
& \times \sum_{n=0}^{\infty} {(-1)^{n} \over n!} 
 \int_{0}^{\beta} d\tau_{1} ... \int_{0}^{\beta} d\tau_{n} \tilde{H}_{A}(\tau_{1}) ... \tilde{H}_{A}(\tau_{n}) \rangle_{0}   \text{  .}
 \end{split}
\end{equation}
To simplify notation, we represent states, both the wave vector and band index, using $K=(\bm{k},s)$ and $\bar{K}=(-\bm{k},s)$. The phonon
states themselves are given by $K_{1}=(\bm{k}+{\bm{q} \over 2},s)$ and $K_{2}=(-\bm{k}+{\bm{q} \over 2},s)$. In other words, we are considering here the product of two propagators which differ in wave vector by a small quantity $\bm{q}$. This leads to beats with
low frequencies $\pm(\omega_{K_{1}}-\omega_{K_{2}})$. The most important contributions come from two modes within the same phonon band $s=s^{\prime}$. Coherence effects are especially strong within the linear-disperse region of a phonon band. Hence, the summation in Eq. \ref{feq3} mixes states with wave vectors $\bm{k}$ and $\bm{k}^{\prime}$, but only within the same phonon band.

Summation of the perturbation expansion is accomplished in two steps. First, a partial summation is affected to obtain dressed propagators. Without considering higher-order terms, this corresponds to the so-called ``dressed-bubble approximation''. Beyond this approximation, we also consider terms in the expansion which couple the two dressed propagators in a ladder expansion. The result is the Bethe-Salpeter equation which must be summed to obtain the thermal response function. We first write for the expansion coefficients of the function $\mathscr{C}_{\bm{q}}(\tau)$,
\begin{equation} 
\mathscr{G}_{\bm{q}} (i \omega_{l}) = {1 \over \beta} \int_{0}^{\beta} \mathscr{C}_{\bm{q}}(\tau) e^{i \hbar \omega_{l} \tau} d\tau
\text{   ,}
\end{equation}
in which $\omega_{l}={2 \pi l \over \beta}$.
Then the correlation function $\mathscr{G}_{\bm{q}}$ is obtained from,
\begin{equation} 
\mathscr{G}_{\bm{q}} (i \omega_{l})  = {1 \over 4 \Omega^{2}} \sum_{K} \sum_{l_{1}} {1 \over 2} \hbar \omega_{K} v_{K \mu} F_{K_{1}K_{2}Q}
(i \omega_{l_{1}}, i \omega_{l}-i \omega_{l_{1}}) 
\text{   .}
\end{equation}
The vertex function $F_{K_{1}K_{2}Q}$ is obtained by summation of ladder diagrams. The lowest order term of the series is 
\begin{equation} \label{f0}
 F_{K_{1}K_{2}Q}^{(0)}
(i \omega_{l_{1}}, i \omega_{l}-i \omega_{l_{1}}) ={1 \over 2} \hbar \omega_{K} v_{K \mu} D_{K_{1}}(i \omega_{l_{1}}) D_{K_{2}} (i \omega_{l}-i \omega_{l_{1}}) 
\text{   .}
\end{equation}
Here $D_{K}(i \omega_{l})$ is the dressed phonon Green's function. 
The vertex function is obtained from iterative solution of the Bethe-Salpeter equation,
\begin{equation} \label{BSE}
\begin{split}
  F_{K_{1}K_{2}Q}  (i \omega_{l_{1}}, i \omega_{l}-i \omega_{l_{1}})= &  F^{(0)}_{K_{1}K_{2}Q } (i \omega_{l_{1}}, i \omega_{l}-i \omega_{l_{1}})    + \\ 36 \beta^{2}    
  & \sum_{K_{3}K_{4}K_{5}} V(\bar{K}_{1} K_{3} K_{5})  V(\bar{K}_{2} K_{4} \bar{K}_{5})     D_{K_{1}}(i \omega_{l_{1}} ) D_{K_{2}}(i \omega_{l}-i \omega_{l_{1}} )     \\
 &   \times \sum_{l_{3}}   
 D_{K_{5}}(i \omega_{l_{1}} -i \omega_{l_{3}} )  F_{K_{3}K_{4}Q} (i \omega_{l_{3}}, i \omega_{l}-i \omega_{l_{3}}   )      
    \end{split}
  \end{equation}                    
                          which is represented in Fig. 2. Following Ref. \cite{Sham:1967aa}, summation on $l_{3}$ is evaluated using contour integration. The only difference 
                         with Ref. \cite{Sham:1967aa} is that we do not use the harmonic approximation of the spectra density associated with the Green's function $ D_{K_{5}}(i \omega_{l_{1}} -i \omega_{l_{3}} ) $. This allows for contributions due to interactions which are not exactly resonant.
                         
                         The ladder diagrams involve products of dressed Green's functions. We analytically continue these products to the neighborhood of the 
real axis, and approximately evaluate the product of the two Green's functions. This involves approximations from Holstein \cite{HOLSTEIN1964410} and was also used by Sham. However, we note that Sham appears to have a sign error in the definition corresponding to Eq. \ref{gfprod} which leads to some disagreement with subsequent expressions \cite{Sham:1967aa,Sham:1967ab}. We find here, for $\eta_{1} $ and $\eta_{2}$ both real and positive,
       \begin{equation} \label{gfprod}
       D_{K_{1}}( \omega^{\prime} + i \eta_{1})D_{K_{2}}(\omega-\omega^{\prime} + i \eta_{2} ) \approx
{2 \pi i \over \beta^{2} \hbar^{2}}    
 \left[{\delta(\omega^{\prime} - \omega_{K}) \over \omega - \bm{q} \cdot \bm{v}_{K} + 2i \Gamma_{K}}
 +{\delta(\omega^{\prime} + \omega_{K}) \over \omega + \bm{q} \cdot \bm{v}_{K} + 2i \Gamma_{K}}
 \right] \text{   .}
    \end{equation}
in which $\bm{v}_{K}$ is the phonon group velocity and
$\Gamma_{K}$ is the phonon linewidth. The expression for the phonon linedwidth can be found in numerous articles including Ref. \cite{Maradudin:1962aa}. 

 The ansatz for $F$ is given, after analytic continuation, by
    \begin{equation}                     
      F_{K_{1}K_{2}Q}  (\omega^{\prime}+ i \eta_{1},  \omega- \omega^{\prime} + i \eta_{2})=X_{QK}(\omega, \omega^{\prime}) \left({2 \pi \over \beta^{2} \hbar^{2}} \right) \left[\delta(\omega^{\prime}-\omega_{K}) +\delta(\omega^{\prime}+\omega_{K})  \right] \text{    .}
        \end{equation}    
        Then for the inhomogeneous term, we have
           \begin{equation}  
           X_{QK}^{(0)}(\omega, \omega^{\prime}) ={1 \over 2} \hbar \omega_{K} v_{K\mu} 
           \left[{\omega_{K} \over 2 \omega_{K} \Gamma_{K} - i \omega \omega_{K} + i \omega^{\prime} \bm{q} \cdot \bm{v}_{K}} \right]
                   \end{equation}          
   This results in the set of coupled linear equations,
        \begin{equation}
        \begin{split}
         n_{K}   \left(n_{K}+1 \right) &  \left(2\Gamma_{K}-i\omega \right) X_{QK}(\omega,\omega_{K}) + 
         n_{K} \left(n_{K}+1 \right)   \left(i\bm{q}  \cdot \bm{v}_{K} \right)  X_{QK}(\omega,\omega_{K}) -\\ 
        {1 \over 2} n_{K} &  \left(n_{K}+1 \right)\hbar \omega_{K} v_{K\mu} = 
          \sum_{K^{\prime}} P_{K K^{\prime}}  X_{QK^{\prime}}(\omega,\omega_{K^{\prime}}) \text{  .}
        \end{split}
                \end{equation}   
                It is also clear from the homogeneous term that $X_{QK}(\omega,-\omega_{K})= -X_{Q\bar{K}}(\omega,\omega_{K})$.  Hence we do not need to write a separate equation for $X_{QK}(\omega,-\omega_{K})$ terms. For simplicity then, we define $Y_{K}(Q,\omega) = X_{QK}(\omega,\omega_{K})$, and then the above equation is written,
                     \begin{equation}
        \begin{split}
         n_{K}   \left(n_{K}+1 \right) &  \left(2\Gamma_{K}-i\omega \right) Y_{K}(Q,\omega)  + 
         n_{K} \left(n_{K}+1 \right)   \left(i\bm{q}  \cdot \bm{v}_{K} \right)  Y_{K}(Q,\omega)  - \\
         {1 \over 2} n_{K} &  \left(n_{K}+1 \right) \hbar \omega_{K} v_{K\mu} = 
          \sum_{K^{\prime}} P_{K K^{\prime}}  Y_{K ^{\prime}}(Q,\omega) \text{  .}
        \end{split}
                \end{equation}     
                In contrast to Sham's work, here we use the spectral density corresponding to a finite phonon linewidth.
                The scattering operator is then
            defined by,
               \begin{equation}
               \begin{split}
                 & \sum_{K^{\prime }}  P_{K K^{\prime}} Y_{K^{\prime}} = 
                 {72 \over \hbar^{2}} \sum_{K^{\prime}K^{\prime \prime}} \left[ |V(K K^{\prime} \bar{K}^{\prime \prime})|^2
                 \left(n_{K}+1\right) \left(n_{K^{\prime}}+1 \right)n_{K^{\prime \prime}}  \right. \\
 & \left. \times \left[ {\Gamma_{K^{\prime }} \over \left(\omega_{K}+\omega_{K^{\prime} }- \omega_{K^{\prime \prime} } \right)^{2} + \Gamma^{2}_{K^{ \prime}}}Y_{K^{\prime \prime}}  
                - {\Gamma_{K^{\prime \prime}} \over \left(\omega_{K}+\omega_{K^{\prime}}- \omega_{K^{\prime \prime}} \right)^{2} + \Gamma^{2}_{K^{\prime \prime}} } Y_{K^{\prime}}
                 \right] +  {1 \over 2} |V(\bar{K} K^{\prime} K^{\prime \prime})|^2\left(n_{K}+1\right) n_{K^{\prime}}n_{K^{\prime \prime}}  \right. \\
 & \left. \times \left[ {\Gamma_{K^{\prime }} \over \left(\omega_{K}-\omega_{K^{\prime }}- \omega_{K^{\prime \prime}} \right)^{2} + \Gamma^{2}_{K^{ \prime}}}Y_{K^{\prime \prime}}  
               + {\Gamma_{K^{\prime \prime}} \over \left(\omega_{K}-\omega_{K^{\prime}}- \omega_{K^{\prime \prime}} \right)^{2} + \Gamma^{2}_{K^{\prime \prime}} } Y_{K^{\prime}}
                 \right]
                 \right]   \text{   .}
                 \end{split}
               \end{equation}  
               Another difference from Ref. \cite{Sham:1967aa} is that the above matrix does not include the diagonal term for $K=K^{\prime}$.
We next apply the transformation of Guyer and Krumhansl \cite{Guyer:1966ab} with
             \begin{equation}
             Y_{K}(Q,\omega) =   Z_{K}(Q,\omega) \left[n_{K}(n_{K}+1) \right]^{-{1 \over 2}}  \text{   ,}
       \end{equation}  
       and also,
          \begin{equation}
          W_{K\mu} ={1 \over 2} \hbar \omega_{K} v_{K\mu} \left[n_{K}(n_{K}+1) \right]^{{1 \over 2}}  \text{   ,}
           \end{equation}          
           and further defining the collision matrix $C$ for $K \ne K^{\prime}$,
               \begin{equation}
               C_{K K^{\prime}} =    \left[n_{K}(n_{K}+1) \right]^{-{1 \over 2}}  P_{K K^{\prime}}   \left[n_{K^{\prime}}(n_{K^{\prime}}+1) \right]^{-{1 \over 2}} \text{   ,}
               \end{equation}  
               then the coupled linear equations are written as,
          \begin{equation}
           \left(2 \Gamma_{K}+i\bm{q}  \cdot \bm{v}_{K}-i\omega \right) Z_{K}(Q,\omega)-
        W_{K\mu} = 
          \sum_{K^{\prime}} C_{K K^{\prime}} Z_{K^{\prime}}(Q,\omega)
                \end{equation}           
 The equation is then written in a convenient form for obtaining iterative solutions,
        \begin{equation} \label{zeq}
    Z_{K}(Q,\omega) =  {i W_{K\mu} +i \sum_{K^{\prime}} C_{K K^{\prime}} Z_{K^{\prime}}(Q,\omega)   \over  \omega - \bm{q}  \cdot \bm{v}_{K} + 2i \Gamma_{K} }
      \end{equation}

The final expression for the imaginary component of response function is computed from
\begin{equation}
\begin{split}
K^{\prime \prime}_{\bm{q}} (\omega) & = 
{c_{V}\beta \hbar \over 2 \Omega^{2}\langle u(\bm{q},0) u(-\bm{q},0) \rangle }
\sum_{K} {1 \over 2} \hbar \omega_{K} v_{K\mu} \left[n_{K}(n_{K}+1) \right]^{1 \over 2} \\
& \times  \left[ \left(n_{B}(\omega)+1 \right)\left(Z_{K}(Q,\omega)+Z^{*}_{K}(Q,\omega)  \right)+
n_{B}(\omega) \left(Z_{K}(Q,-\omega)+Z^{*}_{K}(Q,-\omega)  \right) \right] \text{   .}
\end{split}
\end{equation}
 As usual, the real and imaginary components are connected via the Kramers-Kronig relations. The response function is given in terms of its real and imaginary parts by,
\begin{equation}
K_{\bm{q}} (\omega)  = K^{\prime}_{\bm{q}} (\omega)  +iK^{\prime \prime}_{\bm{q}} (\omega) 
\end{equation}

We  approximate terms related to the energy fluctuations using the classical equipartition theorem. This is partly justified by the assumption that $\bm{q}$ is a small quantity corresponding to long wavelength fluctuations. In addition, $c_{V}=3Nk_{B}$ will also be assumed. Future work will more accurately determine the fluctuation terms. Given these assumptions, we can write for the imaginary part of the response function,
\begin{equation} \label{Kimag}
\begin{split}
K^{\prime \prime}_{\bm{q}} (\omega) & = 
{\beta \hbar \over 2 \Omega k_{B}T^{2} }
\sum_{K} {1 \over 2} \hbar \omega_{K} v_{K\mu} \left[n_{K}(n_{K}+1) \right]^{1 \over 2} \\
& \times  \left[ \left(n_{B}(\omega)+1 \right)\left(Z_{K}(Q,\omega)+Z^{*}_{K}(Q,\omega)  \right)+
n_{B}(\omega) \left(Z_{K}(Q,-\omega)+Z^{*}_{K}(Q,-\omega)  \right) \right] \text{   .}
\end{split}
\end{equation}
It is easily verified that the above equation satisfies the requirement that the imaginary component be an odd function of frequency, $K^{\prime \prime}_{\bm{q}} (-\omega)=-K^{\prime \prime}_{\bm{q}} (\omega)$.

Calculation of the imaginary part of the response function from Eq. \ref{Kimag} and then the real part can be obtained using the Kramers-Kronig relation. Key to the response function calculation
is an iterative algorithm to compute the terms $Z_{K}(Q,\omega)$ from Eq. \ref{zeq}.

\newpage


\end{document}